# EROS variable stars: Fundamental-mode and first-overtone Cepheids in the bar of the Large Magellanic Cloud

**J.P. Beaulieu[1], P. Grison[1], W. Tobin[1,2], J.D. Pritchard[1,2], R. Ferlet[1], F. Lepeintre[1], A. Vidal-Madjar[1], E. Maurice[3], L. Prévot[3], C. Gry[4], J. Guibert[5], O. Moreau[5], F. Tajhamady[5], E. Aubourg[6], P. Bareyre[6], C. Coutures[6], M. Gros[6], B. Laurent[6], M. Lachièze-Rey[6], E. Lesquoy[6], C. Magneville[6], A. Milsztajn[6], L. Moscoso[6], F. Queinnec[6], C. Renault[6], J. Rich[6], M. Spiro[6], L. Vigroux[6], S. Zylberajch[6], R. Ansari[7], F. Cavalier[7], and M. Moniez[7]**

[1] Institut d'Astrophysique de Paris du CNRS, 98bis boulevard Arago, 74014 Paris, France.
[2] Department of Physics and Astronomy, University of Canterbury, Private Bag 4800, Christchurch, New Zealand.
[3] Observatoire de Marseille, 2 place Le Verrier, 13248 Marseille Cedex 04, France.
[4] Laboratoire d'Astronomie Spatiale du CNRS, BP 8, 13376 Marseille Cedex 12, France.
[5] Centre d'Analyse des Images de l'Institut National des Sciences de l'Univers, CNRS, Observatoire de Paris, 61 Avenue de l'Observatoire, 75014 Paris, France.
[6] CEA, DSM/DAPNIA, Centre d'Études de Saclay, 91191 Gif-sur-Yvette, France.
[7] Laboratoire de l'Accélérateur Linéaire IN2P3, Centre d'Orsay, 91405 Orsay, France.

**Abstract.** We present CCD phase-binned light curves at 490 nm for 97 Cepheid variable stars in the bar of the LMC. The photometry was obtained as part of the French EROS project and has excellent phase coverage, permitting accurate decomposition into Fourier components. We identify as 'sinusoidal' or s-Cepheids those stars with periods less than 5.5 d and small second-harmonic components. These stars comprise ~30% of our sample and most form a sequence ~1 mag brighter than the LMC classical Cepheids in the period-luminosity diagram. They are also generally bluer and have lower-amplitude light curves. We infer that the s-Cepheids are first-overtone pulsators because, when their periods are converted to expected fundamental-mode values, they obey a common period-luminosity-colour relation with classical Cepheids. This also confirms the reality of the colour term in the Cepheid period-luminosity-colour relation. Further, the blue edge of the classical Cepheid instability strip agrees well with the theoretical calculations for the fundamental mode made by Chiosi et al. (1993) for the Hertzsprung-Russell and period-luminosity diagrams, but we find that our observed s-Cepheids are $> 0.2$ mag brighter and bluer than the Chiosi et al. predictions for the first-overtone. We identify a number of features in plots of our stars' Fourier-component amplitude ratios and phase differences. These features have been identified with resonances between different pulsation modes. In the LMC we find these features seem to occur at periods very similar to Galactic ones for classical Cepheids, but at different periods for s-Cepheids. We discover a double-mode Cepheid in the LMC, for which $P(\text{first overtone})/P(\text{fundamental}) = 0.710 \pm 0.001$, very similar to observed ratios for Galactic double-mode Cepheids.

*Send offprint requests to*: J.P. Beaulieu, beaulieu@iap.fr

## 1. Introduction

Fourier decomposition techniques are proving to be powerful tools for characterizing the Hertzsprung Progression of the changing form of the light curve as a function of period exhibited by Cepheid variable stars (Hertzsprung 1926), and understanding its physical nature and cause. For single-mode Cepheids, Simon & Lee (1981) adopted a decomposition of the observed magnitudes of form

$$X = X_0 \ + \ \Sigma_{i=1}^{M} X_i \ \cos\,(\ \frac{2\pi}{P}\ i\ (t - t_0) \ + \ \Phi_i) \qquad (1)$$

and were the first to point out the existence of considerable structure in plots of the amplitude ratios

$$R_{k1} \ = \ \frac{X_k}{X_1} \quad k > 1 \qquad (2)$$

and phase differences

$$\Phi_{k1} \ = \ \Phi_k - k\ \Phi_1 \quad k > 1 \qquad (3)$$

against period, $P$, for classical Cepheids in the Galaxy. [The form of Eq. (3) renders $\Phi_{k1}$, which is defined modulo $2\pi$, independent of the epoch, $t_0$.] Subsequent work – Simon (1988), Poretti (1994), and references therein – has shown that with well-determined Fourier components there is a clear separation in these plots between classical Galactic Cepheids with their skew, high-amplitude and possibly bumpy light curves, and another class of Cepheids which were recognized by Hertzsprung (1926) and which Kholopov et al. in the *General Catalogue of Variable Stars* (1985) define as presenting 'light amplitudes below 0.5 mag in $V$ (0.7 in $B$) and almost symmetrical light curves...; as a rule, their periods do not exceed 7 days...' In the past some astronomers have begun refering to these 'symmetrical' or 'sinusoidal' objects as s-Cepheids, and we shall adopt this nomenclature, although a variety of terminologies have been employed.

Around 1958 Payne-Gaposchkin showed that s-Cepheids in the Small Magellanic Cloud are brighter than classical Cepheids of the same period (Gascoigne, 1960; Payne-Gaposchkin 1961; Payne-Gaposchkin & Gaposchkin 1966) and by 1969 this result had been extended to the LMC (Payne-Gaposchkin 1973). It was almost immediately suspected that the s-Cepheids might be overtone pulsators (e.g. Arp 1960).

A first mapping of the Hertzsprung progression for Galactic classical Cepheids in the $R_{21} - P$ and $\Phi_{21} - P$ planes was made by Simon & Lee (1981). On the basis of a few short-period stars these authors, and Gieren (1982), speculated that overtone-mode pulsators might be differently located in the $R_{21} - P$ and $\Phi_{21} - P$ planes. Then Antonello & Poretti (1986), Antonello et al. (1990), Mantegazza & Poretti (1992) and Poretti (1994) determined Fourier coefficients for significant numbers of both Galactic classical and s-Cepheids of short period. They conclude that s-Cepheids follow different sequences in the $R_{21} - P$ and $\Phi_{21} - P$ diagrams. In the $\Phi_{21} - P$ plane there is a rise in $\Phi_{21}$ as periods approach ~3.2 d. At ~3.2 d there is a sharp drop in $\Phi_{21}$, followed by a further rise at longer periods. With a little calligraphic licence we may call this form a 'Z-shape'. The drop of the Z-shape is mirrored by a minimum in $R_{21}$ in the $R_{21} - P$ plane. It was proposed that a separation between classical and s-Cepheids could be based on the $\Phi_{21} - P$ diagram and that the difference should be interpreted as the consequence of different pulsation modes: fundamental radial mode ('1F') for classical Cepheids, and first-overtone radial mode ('1H') for the s-Cepheids. It was further proposed that the s-Cepheid features at $P \sim 3.2$ d should be interpreted as the signature of a resonance between the first and fourth overtone.

On the other hand, Gieren et al. (1990) believe that the overtone nature of Galactic s-Cepheids is established only for $P < 3.2$ d because independent evidence of overtone pulsation has been determined only for some of the stars on this branch of the $\Phi_{21} - P$ diagram. Gieren et al. claim that there is clear evidence against the existence of s-Cepheid overtone pulsations for $P > 3.2$ d.

The differences and similarities between classical and s-Cepheids and their pulsational nature can be clarified by studying stars in the LMC where observed differences in magnitude and colour parallel instrinsic ones. This is because the reddening is relatively slight, even in the bar, where the IRAS maps of Schwering & Israel (1990) indicate an uneven distribution of dust. It is also important to know to what extent the behaviour of Cepheids in the Magellanic Clouds is similar to, or different from, that of Galactic ones, since the study of Cepheids at different metallicities provides new tests of theories of stellar evolution and pulsation. Extragalactic Cepheids are additionally important because they are acting as a cornerstone in establishing the cosmic distance scale.

Andreasen & Petersen (1987) have studied the Fourier components of Cepheids in the Large Magellanic Cloud (LMC). Their observational material was 128 photographic light curves published by Wayman et al. (1984). There were only about 37 points in each $B$ light curve, and 28 in $V$. As a consequence of these limitations there were only 10 s-Cepheids in their sample and the uncertainties on their values of $R_{k1}$ and $\Phi_{k1}$ were large. Within these constraints, the behaviour of LMC and Galactic Cepheids was found to be broadly similar.

In this paper, we improve upon the Andreasen & Petersen results using observations from the EROS project. This project (Expérience de Recherche d'Objets Sombres, Aubourg et al. 1993a,b) has been acquiring photometric time series for large number of stars in the LMC in order to search for possible massive compact objects in the halo of the Galaxy via the brightening due to gravitational microlensing that they should cause of more distant LMC stars. However, EROS CCD observations also provide high-quality light curves for 97 Cepheids in the LMC bar. The high precision and excellent phase coverage (typically 900 points per filter) result in high accuracy Fourier components. Further, the fraction of s-Cepheids in our sample is much larger (~30%), allowing accurate delineation of the classical and s-Cepheids in the amplitude-ratio and other diagrams.

## 2. Observations, reductions and calibrations

In this paper we use the data from the first season of EROS CCD observations in 1991-1992. The EROS CCD equipment has been described fully by Arnaud et al. (1994a,b). The observation, reduction, and calibration procedures have been presented in Grison et al. (1995). To summarize: observations of the LMC bar were obtained at ESO La Silla using a 0.4-m, f/10 reflecting telescope and a $2 \times 8$ mosaic of 16 CCDs. Observations were made in two bandpasses, $B_{\rm E}$ and $R_{\rm E}$, which have mean wavelengths of 490 and 670 nm respectively. Pairs of $B_{\rm E}$ and $R_{\rm E}$ images were obtained approximately every 25 minutes during clear nights between 1991 December and 1992 April. A total of about 1000 images was obtained in each bandpass.

Because large numbers of stars needed to be reduced, a rapid procedure involving the fitting of a Point Spread Function was adopted for the extraction of intrumental magnitudes at the expense of somewhat reduced photometric precision. Zero points were established for the natural EROS magnitudes $B_{\rm E}$ and $R_{\rm E}$ such that a star of zero colour $(B_{\rm E}-R_{\rm E}) \approx 0 \approx (B_{\rm J}-V_{\rm J})$ has its $B_{\rm E}$ magnitude numerically equal to its Johnson $B_{\rm J}$ magnitude and its $R_{\rm E}$ magnitude numerically equal to its Cousins $R_{\rm C}$ magnitude. In this paper we will mostly use the natural magnitudes $B_{\rm E}$ and $R_{\rm E}$ but will also sometimes need to convert to or from Johnson $V_{\rm J}$ and $(B_{\rm J}-V_{\rm J})$ using transformation equations which can be found in Grison et al. (1995):

$$V_{\rm J} = B_{\rm E} - 0.47(B_{\rm E}-R_{\rm E}) \qquad \sigma = 0.07 \text{ mag} \qquad (4)$$

$$(B_{\rm J}-V_{\rm J}) = 0.92(B_{\rm E}-R_{\rm E}) \qquad \sigma = 0.10 \text{ mag} \qquad (5)$$

The EROS photometry has a number of limitations. We will analyze data only from CCDs 0-4 and 7-10 because first-season results from the others were affected by a variety of instrumental or reduction problems. The zero points of the magnitudes may be systematically in error by up to 0.2 mag, and there may be systematic differences of the colour zero points of ∼0.05 mag between different CCDs in the mosaic. Further, crowding or other effects can affect the photometry of individual stars significantly. *In this paper, therefore, it is overall trends which should be considered significant, while discrepancies apparent for individual stars should only be considered suggestive until tested with better-quality photometry with a bigger telescope.*

## 3. Identification of Cepheids and Fourier decomposition

In Figure 1, dots show the colour-magnitude diagram for stars in one CCD of the EROS mosaic. The main sequence and red giant branch are apparent. From the map presented by Page & Carruthers (1981) and the article on which it is based, we estimate the typical reddening in the EROS fields is $E(B_{\rm J}-V_{\rm J})$=0.15 [$\approx E(B_{\rm E}-R_{\rm E})$] with a likely maximum variation of ±0.05. The variations in colour and magnitude due to differential reddening are therefore very similar to the uncertainties in the corresponding photometric zero points and cause little additional smearing in this or other figures.

Variable stars were identified using Grison's (1994) period-searching procedure, which is able to search efficiently for periodic, non-sinusoidal light curves by use of orthogonal combinations of Fourier harmonics. The limits of the Cepheid instability strip were delineated via visual examination of a sample of light curves. To avoid contamination by RR Lyrae stars we only considered variable stars with $B_{\rm E} \leq 17.6$. Then the light curves of all variable stars within this region were examined to exclude eclipsing binaries. The resulting 97 stars form our sample of LMC Cepheids. They are plotted in Figure 1 using symbols other than dots. It can be seen that we have searched to ∼ 1 mag fainter than the lower limit of the region where most Cepheid or Cepheid-like stars occur. Our Cepheids are listed in Table 1. The naming and numbering scheme continues that introduced by Grison et al. (1995). The J2000 equatorial coordinates were derived via Harvard $(x, y)$ coordinates in the manner described by these authors, and are accurate to ∼3 arcsec.

Mean light curves were obtained by sorting the ∼900 best photometric points in each colour into phase order and then taking the mean magnitude averages in bins of 0.01 in phase. Figure 2 presents the $B_{\rm E}$ light curves, which are less noisy than the red ones. Only 29 of our Cepheids were previously identified by Payne-Gaposchkin (1971) on Harvard plates (HV stars). Comparison of our periods with those quoted by Payne-Gaposchkin suggests our typical period uncertainty is about 0.005 days.

Coefficients derived from the Grison period-searching procedure were used to calculate the parameters $X_i$, $\Phi_i$, $R_{k1}$ and $\Phi_{k1}$ for various values of $M$, the order of the Fourier decomposition (defined in Eq. 1). Tests showed that the parameters were only negligibly affected by the value of $M$ (for $M$ greater than 5), except for the few bump Cepheids in our sample (EROS 2081, 2087, 2089...).

We adopted a procedure whereby $M$ was increased until there ceased to be any decrease in the variance $\sigma$ of the residuals between the mean light curves and their Fourier models. Typical values of $\sigma$ are in the range $10^{-2}$ to $5\ 10^{-3}$ mag. Values of $R_{21}$, $R_{31}$, $\Phi_{21}$ and $\Phi_{31}$ for the $B_{\rm E}$ light curves are listed in Table 1. Their uncertainties were calculated as described in the Appendix. Also listed in Table 1 are magnitude means $<B_{\rm E}>$ and $<R_{\rm E}>$ and the peak-to-peak amplitudes $\Delta B_{\rm E}$ and $\Delta R_{\rm E}$.

We can give an estimate of the Cepheid detection efficiency, or, equivalently, the lowest amplitude detectable. These can be derived from the work of Groth (1975) and Scargle (1982). From the statistical properties of the periodogram if a signal is present in the data (Groth 1975), Scargle (1982) has defined a detection efficiency which gives the probability of detecting a periodic signal, $a\cos(\omega t)$, for a given signal to noise ratio, $a/\sigma$, where $a$ is the amplitude of the signal, and $\sigma$ the average precision of the measurements. Applying such results to the EROS data we find with a confidence level better than 99% that we expect to detect all the periodic sinusoidal variable stars with $a/\sigma \simeq 0.3$. The average precision $\sigma$ of the ∼900 CCD measurements has been evaluated from the least-squares residuals of the Cepheids' light curves. Thus, the theoretical detection limit of a sinusoid with a peak-to-peak amplitude $A = 2a$ is typically 0.01 mag for a Cepheid of 14th mag and 0.02 mag for one at 16th mag These predictions are in agreement with the lowest-amplitude contact eclipsing binary found in the EROS data (EROS 1003, $A \simeq 0.04$, $B_{\rm E} = 14.7$; Grison et al. 1995). We have not estimated our detection efficiency for double-mode Cepheids.

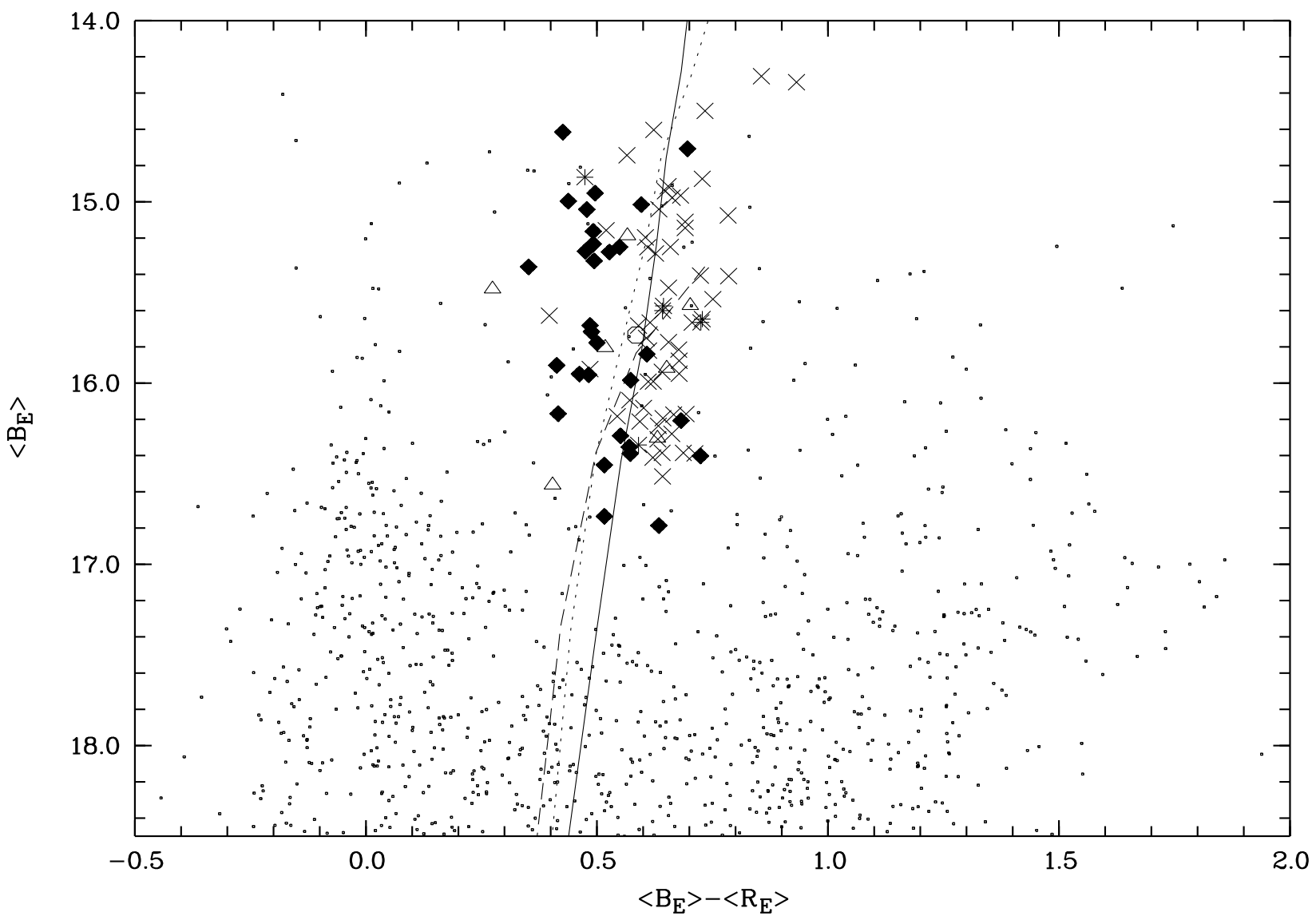

**Fig. 1.** The colour-magnitude diagram of mean values for all stars reduced from CCD 1 for which $< B_{\rm E} >< 18.5$ (dots). The entire sample of Cepheids is also shown: Classical light curves (diagonal crosses), s-Cepheids (filled diamonds), double-mode Cepheid (open circle), intermediate objects in the $R_{21} - P$ plane (asterisks) and anomalous types (open triangles). The same symbols are used in subsequent figures. The full, dotted and dashed lines represent the blue edges of the instability strip for fundamental, first and second overtone pulsation, transformed from the calculations by Chiosi et al. (1993).

## 4. Classification

### *4.1. Anomalous stars and double-mode candidates.*

The distinction between the skew, higher-amplitude light curves of classical Cepheids and the more-sinusoidal, lower amplitude ones of s-Cepheids can be clearly seen in Figure 2. Pairs of (Classical Cepheid, s-Cepheid) with nearly identical periods include (2019, 2020), (2022, 2023), (2034, 2035) and (2066, 2067). However some light curves appear unusual and are likely to confuse the discussion of the separation between classical and s-Cepheids.

EROS 2037 and 2076 have markedly larger scatter in their mean light curves than other stars of similar period. This might be due to photometric contamination by superposed or nearby stars. Another possibility is that the stars might be double-mode Cepheids.

We searched for a second periodicity in these light curves. We subtracted the Grison Fourier model obtained on the mean light curve from the unbinned complete data and then reapplied the Grison algorithm to the residuals. Only EROS 2037 showed a second period, of value $2.443 \pm 0.001$d. For this star the ratio of the second to first periods is then $0.710 \pm 0.001$, in close agreement with the observations reviewed by Balona (1984) for Galactic double-mode Cepheids. This strongly suggests that this star, for which irregular variations were first noted by Shapley & McKibben Nail (1955), is a double-mode Cepheid. It has also been classified by Alcock et al. (1995) as a double-mode Cepheid pulsating in the fundamental-mode and first overtone. The value of the period ratio they obtain is 0.710, like the value we derived.

Andreasen & Petersen (1987) list 7 candidate LMC double-mode Cepheids, but none has been confirmed. Alcock et al. (1995), using data from the MACHO microlensing project, report the detection of 45 double-mode stars amongst 1500 Cepheids. We find only one secure double-mode variable out of 72 stars with $P < 5$d.

The Fourier coefficients for EROS 2026, 2027, 2051 and 2075 may be affected by significant systematic uncertainties because of the gap in the light curves and possible period errors as a consequence of these stars' periods being close to an integral number of days. Some examples of unsatisfactory phase coverage and its effect on Fourier decomposition can be found in Antonello et al. (1990).

Additionally, two of the stars already identified as unusual (2095 and 2097) fall clearly below the general trend in the magnitude-period diagram plotted in Figure 4. Probably they are W Virginis stars associated with the LMC according to the calibration of Demers & Harris (1974). In all the figures in this paper, unusual objects are plotted as open triangles, with the exception of the double-mode star 2037, which is plotted as an open circle.

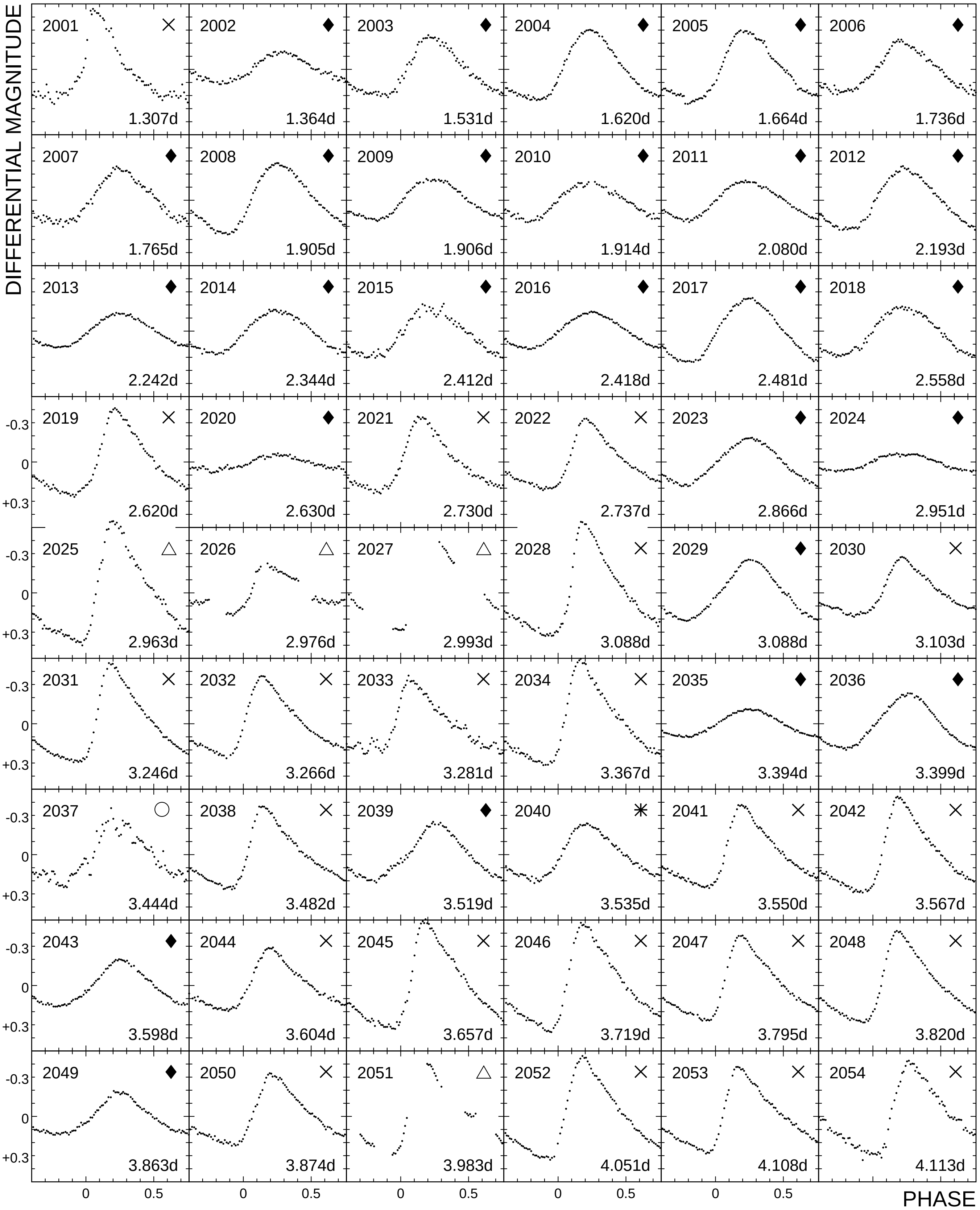

**Fig. 2.** EROS differential $B_{\rm E}$ light curves for 97 Cepheids in the bar of the LMC averaged in 0.01 bins of phase. The value of the magnitude mean $<B_{\rm E}>$ is reported in Table 1. The curves have been phased such that the rising branch attains $<B_{\rm E}>$ at phase zero.

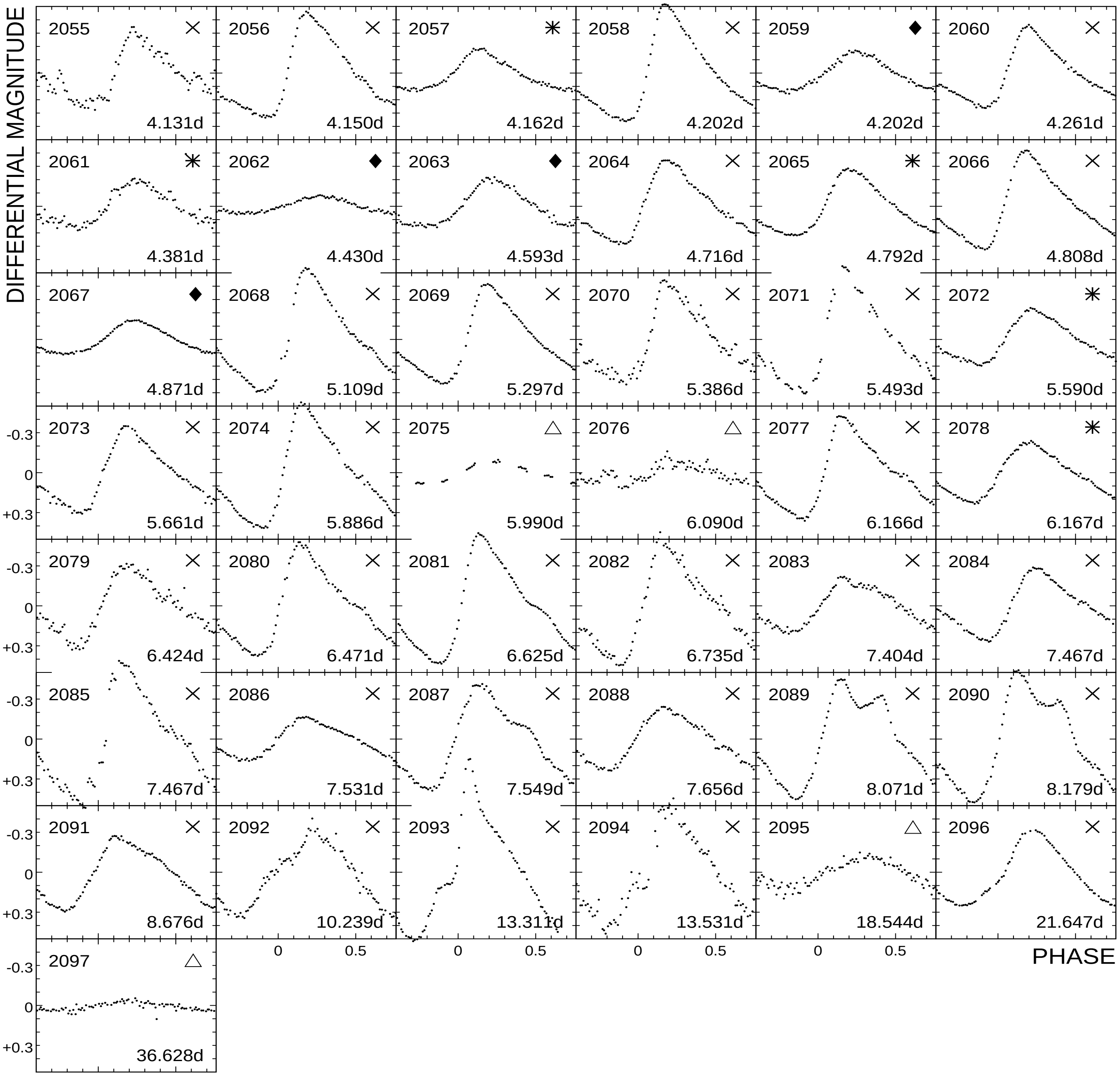

**Fig. 2.** (continued)

**Table 1.** 97 LMC bar Cepheids

| EROS | $\alpha$ | (J2000) | | $P$ | Type | $<B_{\rm E}>$ | $<R_{\rm E}>$ | $R_{21}$ | $R_{31}$ | $\Phi_{21}$ | $\Phi_{31}$ | Remarks |
|---|---|---|---|---|---|---|---|---|---|---|---|---|
| *HV* | $\delta$ | | | day | | $\Delta B_{\rm E}$ | $\Delta R_{\rm E}$ | $\sigma_{R_{21}}$ | $\sigma_{R_{31}}$ | $\sigma_{\Phi_{21}}$ | $\sigma_{\Phi_{31}}$ | |
| 2001 | 5 | 17 | 50.9 | 1.3072 | × | 16.39 | 15.74 | 0.444 | 0.243 | 3.56 | 1.06 | |
| | -69 | 28 | 31 | | | 0.71 | 0.38 | 0.081 | 0.038 | 0.16 | 0.23 | |
| 2002 | 5 | 21 | 4.8 | 1.3637 | ♦ | 16.74 | 16.22 | 0.164 | | 4.04 | | |
| | -69 | 35 | 3 | | | 0.23 | 0.16 | 0.009 | | 0.12 | | |
| 2003 | 5 | 18 | 52.2 | 1.5312 | ♦ | 16.21 | 15.52 | 0.239 | 0.071 | 3.99 | 2.14 | |
| | -69 | 26 | 44 | | | 0.45 | 0.27 | 0.027 | 0.009 | 0.09 | 0.20 | |
| 2004 | 5 | 21 | 20.2 | 1.6205 | ♦ | 16.17 | 15.75 | 0.205 | 0.042 | 3.95 | 2.66 | |
| | -69 | 40 | 51 | | | 0.53 | 0.37 | 0.014 | 0.003 | 0.06 | 0.15 | |
| 2005 | 5 | 16 | 3.2 | 1.6642 | ♦ | 15.90 | 15.49 | 0.212 | 0.055 | 4.15 | 2.22 | |
| | -69 | 28 | 27 | | | 0.54 | 0.35 | 0.005 | 0.017 | 0.07 | 0.18 | |
| 2006 | 5 | 20 | 4.3 | 1.7362 | ♦ | 16.35 | 15.78 | 0.135 | 0.073 | 3.91 | 1.40 | |
| | -69 | 25 | 4 | | | 0.39 | 0.24 | 0.020 | 0.001 | 0.12 | 0.19 | |
| 2007 | 5 | 17 | 10.3 | 1.7652 | ♦ | 16.45 | 15.94 | 0.137 | 0.051 | 3.84 | 1.91 | |
| | -69 | 20 | 55 | | | 0.42 | 0.30 | 0.028 | 0.025 | 0.12 | 0.28 | |
| 2008 | 5 | 17 | 45.4 | 1.9050 | ♦ | 15.95 | 15.49 | 0.210 | 0.036 | 4.44 | 3.42 | |
| | -69 | 35 | 58 | | | 0.53 | 0.35 | 0.013 | 0.004 | 0.06 | 0.18 | |
| 2009 | 5 | 20 | 5.5 | 1.9062 | ♦ | 16.39 | 15.82 | 0.127 | 0.061 | 4.29 | 3.70 | |
| | -69 | 42 | 36 | | | 0.30 | 0.20 | 0.001 | 0.015 | 0.09 | 0.14 | |
| 2010 | 5 | 19 | 32.9 | 1.9139 | ♦ | 16.79 | 16.15 | 0.141 | 0.028 | 5.08 | 5.05 | Faint $<B_{\rm E}>$ |
| | -69 | 36 | 32 | | | 0.29 | 0.20 | 0.029 | 0.028 | 0.11 | 0.52 | 1F pulsator ? |
| 2011 | 5 | 22 | 31.5 | 2.0801 | ♦ | 16.29 | 15.74 | 0.119 | | 4.93 | | |
| | -69 | 39 | 52 | | | 0.30 | 0.21 | 0.014 | | 0.07 | | |
| 2012 | 5 | 19 | 32.4 | 2.1925 | ♦ | 15.95 | 15.47 | 0.141 | 0.023 | 4.38 | 2.97 | |
| | -69 | 29 | 11 | | | 0.46 | 0.30 | 0.002 | 0.013 | 0.07 | 0.35 | |
| 2013 | 5 | 17 | 43.7 | 2.2422 | ♦ | 15.84 | 15.23 | 0.086 | | 4.31 | | |
| | -69 | 37 | 39 | | | 0.26 | 0.17 | 0.006 | | 0.09 | | |
| 2014 | 5 | 18 | 34.2 | 2.3442 | ♦ | 15.78 | 15.28 | 0.064 | 0.020 | 4.91 | 3.48 | |
| | -69 | 32 | 15 | | | 0.33 | 0.23 | 0.009 | 0.005 | 0.15 | 0.46 | |
| 2015 | 5 | 18 | 27.6 | 2.4121 | ♦ | 15.00 | 14.56 | 0.130 | 0.039 | 4.56 | 2.44 | Optical Binary ? |
| | -69 | 19 | 30 | | | 0.37 | 0.26 | 0.024 | 0.040 | 0.16 | 0.53 | |
| 2016 | 5 | 17 | 43.1 | 2.4184 | ♦ | 15.72 | 15.23 | 0.063 | 0.013 | 4.56 | 5.59 | |
| | -69 | 30 | 32 | | | 0.28 | 0.18 | 0.010 | 0.007 | 0.11 | 0.49 | |
| 2017 | 5 | 17 | 41.3 | 2.4806 | ♦ | 15.68 | 15.20 | 0.097 | 0.019 | 4.51 | 4.44 | |
| | -69 | 32 | 42 | | | 0.49 | 0.33 | 0.001 | 0.009 | 0.07 | 0.30 | |
| 2018 | 5 | 16 | 10.9 | 2.5582 | ♦ | 15.98 | 15.41 | 0.058 | 0.045 | 4.85 | 2.45 | |
| | -69 | 18 | 5 | | | 0.36 | 0.24 | 0.023 | 0.011 | 0.21 | 0.24 | |
| 2019 | 5 | 19 | 41.9 | 2.6196 | × | 16.41 | 15.79 | 0.353 | 0.130 | 3.96 | 1.81 | |
| | -69 | 23 | 52 | | | 0.66 | 0.42 | 0.038 | 0.023 | 0.11 | 0.16 | |
| 2020 | 5 | 21 | 40.6 | 2.6303 | ♦ | 15.36 | 15.01 | 0.120 | 0.053 | 3.83 | 0.78 | |
| | -69 | 35 | 50 | | | 0.14 | 0.08 | 0.045 | 0.019 | 0.27 | 0.61 | |
| 2021 | 5 | 19 | 2.9 | 2.7295 | × | 16.21 | 15.62 | 0.374 | 0.129 | 4.02 | 1.80 | |
| | -69 | 21 | 20 | | | 0.57 | 0.37 | 0.018 | 0.022 | 0.11 | 0.17 | |
| 2022 | 5 | 27 | 17.2 | 2.7374 | × | 16.31 | 15.68 | 0.402 | 0.166 | 4.07 | 1.95 | |
| | -69 | 44 | 21 | | | 0.53 | 0.36 | 0.042 | 0.013 | 0.12 | 0.17 | |
| 2023 | 5 | 16 | 39.5 | 2.8661 | ♦ | 16.40 | 15.68 | 0.023 | 0.034 | 3.53 | 5.64 | Faint $<B_{\rm E}>$ |
| | -69 | 21 | 45 | | | 0.36 | 0.23 | 0.007 | 0.004 | 0.36 | 0.21 | |
| 2024 | 5 | 20 | 19.8 | 2.9506 | ♦ | 15.27 | 14.80 | 0.030 | 0.085 | 4.87 | 3.11 | |
| | -69 | 26 | 48 | | | 0.12 | 0.08 | 0.015 | 0.029 | 0.59 | 0.18 | |
| 2025 | 5 | 18 | 50.0 | 2.9629 | △ | 15.80 | 15.29 | 0.392 | 0.213 | 4.13 | 2.06 | |
| | -69 | 21 | 32 | | | 0.94 | 0.65 | 0.053 | 0.010 | 0.12 | 0.16 | |
| 2026 | 5 | 15 | 37.1 | 2.9755 | △ | 15.48 | 15.21 | 0.456 | 0.522 | 3.78 | 2.87 | |
| | -69 | 30 | 29 | | | 0.39 | 0.31 | 0.151 | 0.015 | 0.24 | 0.29 | |
| 2027 | 5 | 21 | 59.6 | 2.9933 | △ | 16.30 | 15.67 | 0.309 | 0.165 | 4.31 | 2.19 | Classical Cepheid ? |
| | -69 | 43 | 4 | | | 0.66 | 0.51 | 0.023 | 0.013 | 0.16 | 0.21 | |
| 2028 | 5 | 26 | 28.7 | 3.0876 | × | 15.92 | 15.44 | 0.428 | 0.220 | 4.07 | 2.03 | |
| | -69 | 51 | 16 | | | 0.87 | 0.64 | 0.058 | 0.022 | 0.14 | 0.19 | |

**Table 1.** 97 LMC bar Cepheids

| EROS *HV* | $\alpha$ $\delta$ (J2000) | | | $P$ day | Type | $<B_{\rm E}>$ $\Delta B_{\rm E}$ | $<R_{\rm E}>$ $\Delta R_{\rm E}$ | $R_{21}$ $\sigma_{R_{21}}$ | $R_{31}$ $\sigma_{R_{31}}$ | $\Phi_{21}$ $\sigma_{\Phi_{21}}$ | $\Phi_{31}$ $\sigma_{\Phi_{31}}$ | Remarks |
|---|---|---|---|---|---|---|---|---|---|---|---|---|
| 2029 | 5 | 19 | 44.6 | 3.0882 | ♦ | 15.33 | 14.83 | 0.083 | 0.048 | 3.38 | 0.08 | |
| *5757* | -69 | 22 | 56 | | | 0.46 | 0.29 | 0.016 | 0.003 | 0.10 | 0.13 | |
| 2030 | 5 | 19 | 38.9 | 3.1030 | × | 16.24 | 15.60 | 0.328 | 0.114 | 4.05 | 1.92 | |
| | -69 | 22 | 3 | | | 0.44 | 0.30 | 0.031 | 0.003 | 0.10 | 0.15 | |
| 2031 | 5 | 15 | 49.7 | 3.2456 | × | 16.09 | 15.52 | 0.397 | 0.208 | 4.13 | 2.08 | |
| | -69 | 30 | 28 | | | 0.75 | 0.48 | 0.053 | 0.022 | 0.12 | 0.17 | |
| 2032 | 5 | 18 | 39.7 | 3.2664 | × | 16.28 | 15.62 | 0.404 | 0.173 | 4.14 | 2.28 | |
| | -69 | 35 | 15 | | | 0.61 | 0.40 | 0.015 | 0.003 | 0.12 | 0.16 | |
| 2033 | 5 | 17 | 57.2 | 3.2812 | × | 15.63 | 15.23 | 0.369 | 0.226 | 4.09 | 1.74 | Optical Binary ? |
| | -69 | 34 | 52 | | | 0.53 | 0.42 | 0.014 | 0.021 | 0.14 | 0.19 | |
| 2034 | 5 | 17 | 42.4 | 3.3674 | × | 16.39 | 15.68 | 0.402 | 0.210 | 4.21 | 2.19 | |
| | -69 | 21 | 33 | | | 0.80 | 0.55 | 0.052 | 0.024 | 0.13 | 0.17 | |
| 2035 | 5 | 18 | 2.0 | 3.3943 | ♦ | 15.25 | 14.70 | 0.051 | | 3.73 | | |
| | -69 | 37 | 14 | | | 0.21 | 0.14 | 0.013 | | 0.14 | | |
| 2036 | 5 | 21 | 52.5 | 3.3990 | ♦ | 15.28 | 14.75 | 0.082 | 0.030 | 3.34 | 5.20 | |
| | -69 | 39 | 55 | | | 0.41 | 0.29 | 0.010 | 0.007 | 0.08 | 0.18 | |
| 2037 | 5 | 27 | 15.6 | 3.4438 | ○ | 15.73 | 15.15 | 0.217 | 0.073 | 4.08 | 1.92 | Double-mode |
| *970* | -69 | 43 | 40 | | | 0.59 | 0.39 | 0.064 | 0.071 | 0.19 | 0.50 | |
| 2038 | 5 | 18 | 29.7 | 3.4822 | × | 16.51 | 15.87 | 0.439 | 0.196 | 4.22 | 2.25 | |
| | -69 | 34 | 0 | | | 0.63 | 0.46 | 0.013 | 0.025 | 0.13 | 0.18 | |
| 2039 | 5 | 16 | 29.4 | 3.5186 | ♦ | 15.23 | 14.74 | 0.121 | 0.081 | 3.26 | 0.26 | |
| | -69 | 32 | 7 | | | 0.44 | 0.28 | 0.010 | 0.005 | 0.11 | 0.13 | |
| 2040 | 5 | 19 | 41.0 | 3.5347 | ∗ | 16.34 | 15.75 | 0.237 | 0.055 | 4.22 | 2.36 | 1F pulsator |
| | -69 | 34 | 17 | | | 0.43 | 0.30 | 0.021 | 0.017 | 0.07 | 0.18 | |
| 2041 | 5 | 26 | 35.9 | 3.5496 | × | 15.99 | 15.37 | 0.406 | 0.190 | 4.15 | 2.19 | |
| *12043* | -69 | 49 | 14 | | | 0.63 | 0.43 | 0.014 | 0.024 | 0.12 | 0.17 | |
| 2042 | 5 | 17 | 1.5 | 3.5668 | × | 16.18 | 15.64 | 0.397 | 0.184 | 4.15 | 2.13 | |
| | -69 | 30 | 35 | | | 0.72 | 0.95 | 0.042 | 0.016 | 0.12 | 0.16 | |
| 2043 | 5 | 16 | 7.3 | 3.5982 | ♦ | 15.16 | 14.67 | 0.130 | 0.049 | 3.52 | 0.70 | |
| | -69 | 20 | 48 | | | 0.35 | 0.23 | 0.014 | 0.013 | 0.08 | 0.15 | |
| 2044 | 5 | 17 | 0.3 | 3.6042 | × | 16.39 | 15.70 | 0.342 | 0.115 | 4.14 | 2.01 | |
| | -69 | 20 | 7 | | | 0.47 | 0.31 | 0.028 | 0.017 | 0.10 | 0.15 | |
| 2045 | 5 | 22 | 48.0 | 3.6565 | × | 15.99 | 15.38 | 0.372 | 0.190 | 4.18 | 2.24 | |
| *2478* | -69 | 42 | 41 | | | 0.82 | 0.54 | 0.049 | 0.031 | 0.12 | 0.17 | |
| 2046 | 5 | 18 | 57.2 | 3.7186 | × | 16.19 | 15.55 | 0.401 | 0.182 | 4.22 | 2.40 | |
| *5752* | -69 | 34 | 1 | | | 0.82 | 0.55 | 0.037 | 0.015 | 0.12 | 0.17 | |
| 2047 | 5 | 18 | 7.3 | 3.7953 | × | 16.17 | 15.48 | 0.397 | 0.186 | 4.18 | 2.33 | |
| *12008* | -69 | 38 | 59 | | | 0.65 | 0.44 | 0.041 | 0.017 | 0.13 | 0.17 | |
| 2048 | 5 | 22 | 40.3 | 3.8202 | × | 15.94 | 15.30 | 0.396 | 0.167 | 4.23 | 2.21 | |
| | -69 | 40 | 24 | | | 0.69 | 0.46 | 0.042 | 0.022 | 0.12 | 0.17 | |
| 2049 | 5 | 16 | 50.9 | 3.8626 | ♦ | 15.04 | 14.56 | 0.183 | 0.056 | 3.55 | 0.68 | |
| | -69 | 19 | 31 | | | 0.32 | 0.23 | 0.004 | 0.018 | 0.08 | 0.19 | |
| 2050 | 5 | 21 | 52.6 | 3.8735 | × | 16.17 | 15.51 | 0.351 | 0.134 | 4.14 | 2.16 | |
| | -69 | 36 | 19 | | | 0.54 | 0.36 | 0.003 | 0.014 | 0.10 | 0.14 | |
| 2051 | 5 | 16 | 56.9 | 3.9828 | △ | 15.92 | 15.27 | 0.539 | 0.232 | 4.48 | 2.03 | Classical Cepheid ? |
| | -69 | 39 | 19 | | | 0.71 | 0.45 | 0.090 | 0.036 | 0.19 | 0.29 | |
| 2052 | 5 | 26 | 18.6 | 4.0506 | × | 15.82 | 15.21 | 0.379 | 0.166 | 4.24 | 2.38 | |
| *961* | -69 | 48 | 4 | | | 0.78 | 0.52 | 0.001 | 0.012 | 0.11 | 0.15 | |
| 2053 | 5 | 17 | 55.7 | 4.1083 | × | 15.75 | 15.14 | 0.424 | 0.208 | 4.31 | 2.38 | |
| | -69 | 34 | 54 | | | 0.66 | 0.45 | 0.047 | 0.024 | 0.13 | 0.18 | |
| 2054 | 5 | 20 | 18.0 | 4.1128 | × | 15.88 | 15.20 | 0.404 | 0.155 | 4.27 | 2.44 | |
| | -69 | 30 | 13 | | | 0.70 | 0.43 | 0.041 | 0.028 | 0.12 | 0.18 | |
| 2055 | 5 | 26 | 31.5 | 4.1306 | × | 15.16 | 14.64 | 0.484 | 0.119 | 4.33 | 2.63 | Optical binary ? |
| | -69 | 50 | 23 | | | 0.57 | 0.41 | 0.066 | 0.063 | 0.15 | 0.34 | |
| 2056 | 5 | 26 | 28.5 | 4.1501 | × | 15.68 | 15.09 | 0.363 | 0.169 | 4.18 | 2.41 | |
| *962* | -69 | 44 | 14 | | | 0.78 | 0.52 | 0.046 | 0.024 | 0.11 | 0.16 | |

**Table 1.** 97 LMC bar Cepheids

| EROS<br>*HV* | $\alpha$<br>$\delta$ (J2000) | | | $P$<br>day | Type | $<B_E>$<br>$\Delta B_E$ | $<R_E>$<br>$\Delta R_E$ | $R_{21}$<br>$\sigma_{R_{21}}$ | $R_{31}$<br>$\sigma_{R_{31}}$ | $\Phi_{21}$<br>$\sigma_{\Phi_{21}}$ | $\Phi_{31}$<br>$\sigma_{\Phi_{31}}$ | Remarks |
|---|---|---|---|---|---|---|---|---|---|---|---|---|
| 2057 | 5 | 26 | 48.7 | 4.1619 | $\ast$ | 14.86 | 14.39 | 0.249 | 0.088 | 3.79 | 1.22 | 1F pulsator |
| | -69 | 51 | 31 | | | 0.31 | 0.22 | 0.028 | 0.007 | 0.09 | 0.17 | |
| 2058 | 5 | 18 | 51.8 | 4.2024 | $\times$ | 15.77 | 15.12 | 0.404 | 0.181 | 4.24 | 2.31 | |
| *12010* | -69 | 39 | 11 | | | 0.88 | 0.58 | 0.028 | 0.001 | 0.12 | 0.17 | |
| 2059 | 5 | 16 | 25.8 | 4.2024 | $\blacklozenge$ | 14.95 | 14.46 | 0.129 | 0.043 | 3.78 | 0.40 | |
| | -69 | 29 | 53 | | | 0.30 | 0.20 | 0.003 | 0.025 | 0.11 | 0.30 | |
| 2060 | 5 | 21 | 5.3 | 4.2615 | $\times$ | 16.14 | 15.54 | 0.405 | 0.153 | 4.29 | 2.44 | |
| | -69 | 40 | 32 | | | 0.61 | 0.44 | 0.037 | 0.006 | 0.11 | 0.16 | |
| 2061 | 5 | 18 | 29.1 | 4.3812 | $\ast$ | 15.60 | 14.96 | 0.221 | 0.098 | 4.62 | 2.21 | |
| | -69 | 27 | 48 | | | 0.35 | 0.23 | 0.089 | 0.099 | 0.27 | 0.55 | |
| 2062 | 5 | 17 | 27.1 | 4.4302 | $\blacklozenge$ | 14.61 | 14.19 | 0.107 | | 3.16 | | |
| | -69 | 20 | 8 | | | 0.13 | 0.09 | 0.011 | | 0.18 | | |
| 2063 | 5 | 27 | 32.8 | 4.5929 | $\blacklozenge$ | 15.02 | 14.42 | 0.157 | 0.067 | 3.90 | 1.68 | |
| | -69 | 49 | 12 | | | 0.35 | 0.23 | 0.029 | 0.025 | 0.10 | 0.20 | |
| 2064 | 5 | 16 | 53.2 | 4.7162 | $\times$ | 15.95 | 15.27 | 0.363 | 0.116 | 4.39 | 2.60 | |
| *2422* | -69 | 22 | 5 | | | 0.63 | 0.41 | 0.031 | 0.010 | 0.10 | 0.15 | |
| 2065 | 5 | 17 | 18.7 | 4.7923 | $\ast$ | 15.57 | 14.93 | 0.264 | 0.080 | 4.17 | 2.05 | 1F pulsator |
| *12006* | -69 | 32 | 59 | | | 0.49 | 0.34 | 0.013 | 0.001 | 0.08 | 0.12 | |
| 2066 | 5 | 17 | 57.2 | 4.8079 | $\times$ | 15.81 | 15.14 | 0.404 | 0.152 | 4.40 | 2.57 | |
| *2439* | -69 | 38 | 52 | | | 0.74 | 0.50 | 0.036 | 0.013 | 0.11 | 0.15 | |
| 2067 | 5 | 17 | 2.6 | 4.8713 | $\blacklozenge$ | 14.71 | 14.01 | 0.154 | 0.047 | 3.84 | 1.50 | |
| | -69 | 38 | 51 | | | 0.25 | 0.15 | 0.015 | 0.005 | 0.06 | 0.14 | |
| 2068 | 5 | 19 | 18.2 | 5.1094 | $\times$ | 15.67 | 15.05 | 0.420 | 0.152 | 4.38 | 2.31 | |
| *5749* | -69 | 30 | 27 | | | 0.91 | 0.61 | 0.032 | 0.005 | 0.12 | 0.18 | |
| 2069 | 5 | 26 | 17.6 | 5.2965 | $\times$ | 15.47 | 14.82 | 0.382 | 0.141 | 4.35 | 2.51 | |
| *2531* | -69 | 48 | 29 | | | 0.75 | 0.50 | 0.031 | 0.001 | 0.10 | 0.15 | |
| 2070 | 5 | 17 | 32.6 | 5.3862 | $\times$ | 14.60 | 13.98 | 0.376 | 0.162 | 4.22 | 2.38 | Optical binary ? |
| *2430* | -69 | 25 | 14 | | | 0.75 | 0.47 | 0.036 | 0.041 | 0.13 | 0.22 | |
| 2071 | 5 | 19 | 3.3 | 5.4926 | $\times$ | 15.25 | 14.64 | 0.420 | 0.168 | 4.37 | 2.49 | |
| | -69 | 40 | 8 | | | 0.93 | 0.63 | 0.043 | 0.032 | 0.14 | 0.21 | |
| 2072 | 5 | 19 | 16.0 | 5.5896 | $\ast$ | 15.65 | 14.92 | 0.261 | 0.066 | 4.35 | 2.82 | 1F pulsator |
| *12011* | -69 | 41 | 26 | | | 0.41 | 0.28 | 0.010 | 0.008 | 0.08 | 0.15 | |
| 2073 | 5 | 19 | 38.7 | 5.6609 | $\times$ | 15.67 | 14.96 | 0.357 | 0.103 | 4.46 | 2.62 | |
| *2455* | -69 | 37 | 42 | | | 0.66 | 0.44 | 0.019 | 0.014 | 0.10 | 0.16 | |
| 2074 | 5 | 26 | 59.5 | 5.8858 | $\times$ | 15.20 | 14.59 | 0.413 | 0.148 | 4.48 | 2.59 | |
| *965* | -69 | 51 | 7 | | | 0.93 | 0.63 | 0.033 | 0.024 | 0.12 | 0.18 | |
| 2075 | 5 | 26 | 44.3 | 5.9902 | $\triangle$ | 15.57 | 14.87 | 0.171 | | 4.60 | | |
| | -69 | 48 | 1 | | | 0.17 | 0.12 | 0.033 | | 0.13 | | |
| 2076 | 5 | 26 | 47.9 | 6.0899 | $\triangle$ | 13.97 | 13.20 | 0.413 | 0.415 | 3.58 | 2.72 | Larger scatter |
| | -69 | 50 | 18 | | | 0.17 | 0.08 | 0.162 | 0.127 | 0.30 | 0.35 | |
| 2077 | 5 | 16 | 6.3 | 6.1662 | $\times$ | 15.07 | 14.29 | 0.436 | 0.169 | 4.54 | 2.51 | |
| *922* | -69 | 28 | 24 | | | 0.78 | 0.46 | 0.038 | 0.026 | 0.12 | 0.18 | |
| 2078 | 5 | 16 | 59.3 | 6.1668 | $\ast$ | 15.66 | 14.94 | 0.270 | 0.028 | 4.61 | 3.21 | 1F pulsator |
| | -69 | 23 | 40 | | | 0.46 | 0.30 | 0.008 | 0.013 | 0.07 | 0.27 | |
| 2079 | 5 | 15 | 52.2 | 6.4236 | $\times$ | 15.28 | 14.66 | 0.384 | 0.074 | 4.70 | 3.17 | |
| | -69 | 28 | 15 | | | 0.62 | 0.40 | 0.047 | 0.002 | 0.12 | 0.33 | |
| 2080 | 5 | 18 | 42.6 | 6.4708 | $\times$ | 15.54 | 14.79 | 0.420 | 0.159 | 4.58 | 2.55 | |
| *12009* | -69 | 38 | 21 | | | 0.84 | 0.54 | 0.043 | 0.010 | 0.12 | 0.18 | |
| 2081 | 5 | 22 | 19.2 | 6.6254 | $\times$ | 15.25 | 14.59 | 0.423 | 0.148 | 4.53 | 2.56 | |
| *944* | -69 | 37 | 51 | | | 0.97 | 0.64 | 0.040 | 0.006 | 0.12 | 0.18 | |
| 2082 | 5 | 18 | 4.7 | 6.7348 | $\times$ | 14.91 | 14.26 | 0.396 | 0.144 | 4.71 | 2.63 | |
| *2438* | -69 | 25 | 38 | | | 0.91 | 0.64 | 0.024 | 0.011 | 0.11 | 0.18 | |
| 2083 | 5 | 27 | 23.2 | 7.4044 | $\times$ | 15.41 | 14.68 | 0.207 | 0.078 | 4.98 | 2.98 | |
| | -69 | 50 | 55 | | | 0.41 | 0.26 | 0.025 | 0.005 | 0.10 | 0.22 | |

**Table 1.** 97 LMC bar Cepheids

| EROS *HV* | $\alpha$ $\delta$ (J2000) | | | $P$ day | Type | $<B_\mathrm{E}>$ $\Delta B_\mathrm{E}$ | $<R_\mathrm{E}>$ $\Delta R_\mathrm{E}$ | $R_{21}$ $\sigma_{R_{21}}$ | $R_{31}$ $\sigma_{R_{31}}$ | $\Phi_{21}$ $\sigma_{\Phi_{21}}$ | $\Phi_{31}$ $\sigma_{\Phi_{31}}$ | Remarks |
|---|---|---|---|---|---|---|---|---|---|---|---|---|
| 2084 | 5 | 26 | 49.6 | 7.4674 | $\times$ | 15.11 | 14.42 | 0.354 | 0.076 | 4.62 | 2.71 | |
| | -69 | 45 | 49 | | | 0.54 | 0.37 | 0.005 | 0.010 | 0.09 | 0.16 | |
| 2085 | 5 | 27 | 34.4 | 7.4674 | $\times$ | 14.94 | 14.29 | 0.422 | 0.172 | 4.61 | 2.38 | |
| | -69 | 51 | 20 | | | 1.07 | 0.69 | 0.026 | 0.041 | 0.13 | 0.21 | |
| 2086 | 5 | 26 | 44.4 | 7.5314 | $\times$ | 15.04 | 14.41 | 0.238 | 0.049 | 4.83 | 2.23 | |
| | -69 | 48 | 6 | | | 0.32 | 0.24 | 0.021 | 0.016 | 0.07 | 0.21 | |
| 2087 | 5 | 16 | 55.3 | 7.5492 | $\times$ | 15.14 | 14.45 | 0.321 | 0.122 | 4.80 | 2.47 | |
| *927* | -69 | 19 | 52 | | | 0.79 | 0.50 | 0.015 | 0.016 | 0.09 | 0.14 | |
| 2088 | 5 | 16 | 19.6 | 7.6565 | $\times$ | 15.41 | 14.63 | 0.230 | 0.039 | 4.73 | 3.51 | |
| | -69 | 18 | 21 | | | 0.48 | 0.28 | 0.023 | 0.014 | 0.08 | 0.32 | |
| 2089 | 5 | 27 | 5.0 | 8.0714 | $\times$ | 14.98 | 14.31 | 0.256 | 0.212 | 5.14 | 3.54 | |
| *968* | -69 | 50 | 41 | | | 0.91 | 0.60 | 0.042 | 0.027 | 0.15 | 0.14 | |
| 2090 | 5 | 17 | 25.3 | 8.1795 | $\times$ | 14.74 | 14.18 | 0.276 | 0.220 | 4.87 | 3.15 | |
| *2426* | -69 | 20 | 59 | | | 0.99 | 0.67 | 0.046 | 0.010 | 0.13 | 0.14 | |
| 2091 | 5 | 16 | 0.2 | 8.6761 | $\times$ | 14.97 | 14.28 | 0.251 | 0.056 | 5.01 | 2.91 | |
| *923* | -69 | 32 | 17 | | | 0.56 | 0.36 | 0.012 | 0.016 | 0.07 | 0.18 | |
| 2092 | 5 | 16 | 9.9 | 10.2390 | $\times$ | 14.87 | 14.15 | 0.080 | 0.114 | 0.82 | 5.98 | |
| *2414* | -69 | 32 | 39 | | | 0.67 | 0.38 | 0.044 | 0.038 | 0.29 | 0.18 | |
| 2093 | 5 | 19 | 15.5 | 13.3110 | $\times$ | 14.50 | 13.76 | 0.244 | 0.161 | 4.38 | 1.30 | |
| *932* | -69 | 36 | 16 | | | 1.36 | 0.88 | 0.017 | 0.030 | 0.18 | 0.25 | |
| 2094 | 5 | 18 | 23.0 | 13.5310 | $\times$ | 14.31 | 13.45 | 0.174 | 0.113 | 4.19 | 0.96 | |
| | -69 | 21 | 50 | | | 0.92 | 0.61 | 0.042 | 0.021 | 0.21 | 0.29 | |
| 2095 | 5 | 21 | 16.3 | 18.5440 | $\triangle$ | 16.56 | 16.16 | 0.077 | | 0.42 | | Below PL relation |
| | -69 | 38 | 31 | | | 0.24 | 0.21 | 0.065 | | 0.50 | | |
| 2096 | 5 | 19 | 28.7 | 21.6470 | $\times$ | 14.34 | 13.41 | 0.149 | 0.081 | 3.74 | 1.04 | |
| *2453* | -69 | 30 | 32 | | | 0.56 | 0.36 | 0.019 | 0.010 | 0.09 | 0.10 | |
| 2097 | 5 | 20 | 32.6 | 36.6280 | $\triangle$ | 15.19 | 14.62 | 0.276 | | 4.38 | | Below PL relation |
| | -69 | 42 | 20 | | | 0.07 | 0.07 | 0.145 | | 0.41 | | |

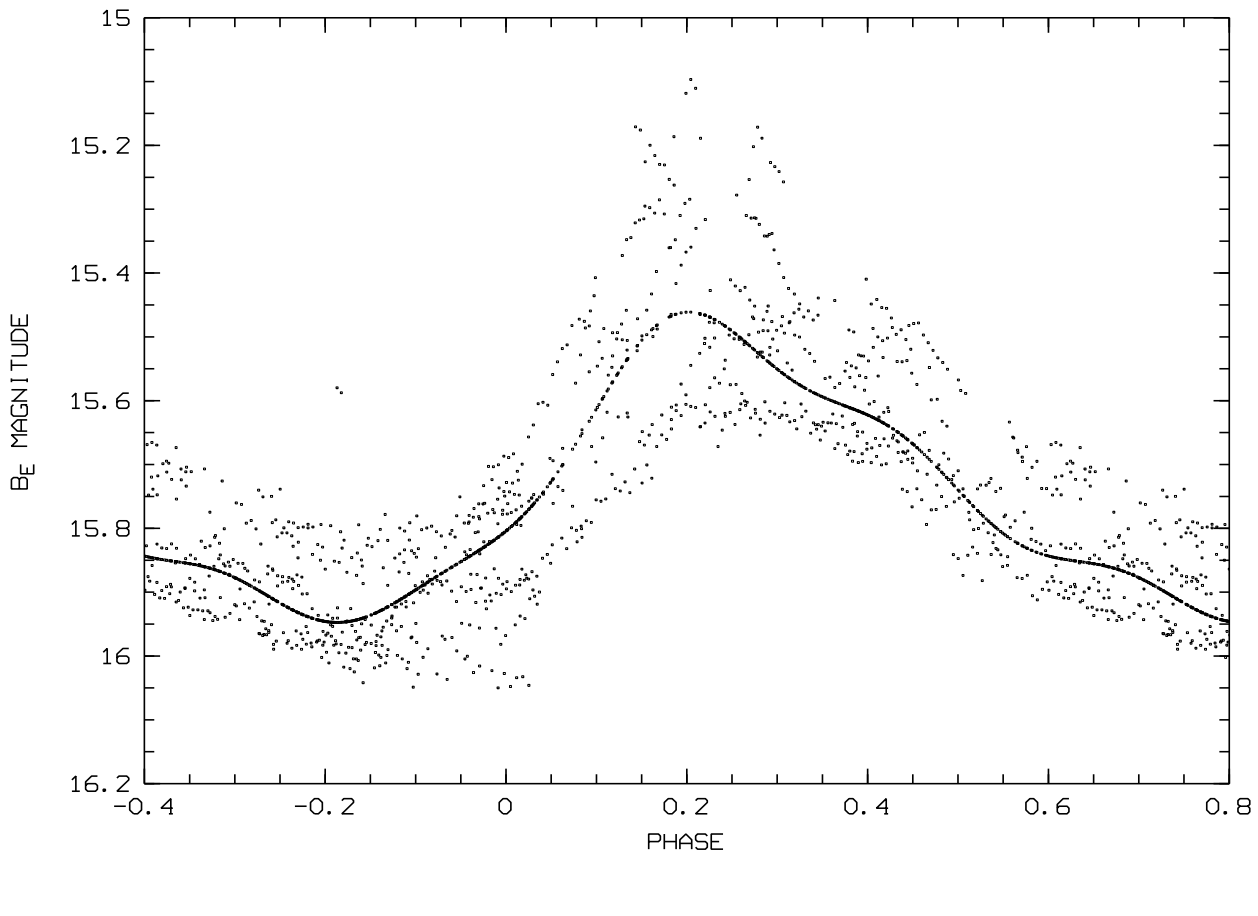

**Fig. 3.** (a) The fundamental mode for EROS 2037 ($P$ = 3.44 d). The points show the full, unbinned $B_\mathrm{E}$ light curve. The full line shows the Fourier model to the phase-binned data of Fig. 2.

### *4.2. Classical, sinusoidal and indeterminate Cepheids*

Figures 5 and 6 show plots of $R_{21}$ and $\Phi_{21}$ against period. It is evident that the points fall in several groups, and as noted above, earlier authors have adopted various classification criteria to distinguish classical from s-Cepheids.

It seems to us that the most transparent discrimination on the form of the light curve should be in the $R_{21} - P$ plane where a low value of $R_{21}$ indicates more sinusoidal light curves. Mindful of earlier work on Galactic Cepheids, we have selected as s-Cepheids those stars lying in the region defined by $R_{21} < 0.30$ for $P < 3$ d and $R_{21} < 0.20$ for $3.0 < P < 5.5$ d. [This last upper limit is not well-defined by our data. Work on Galactic Cepheids (e.g. Simon & Lee 1981) shows that the $R_{21}$ value for classical Cepheids - as defined by the period-luminosity relation - falls to low values for $P \approx 7 - 10$ d.] Our s-Cepheids are plotted in all figures as filled diamonds. Our sample is composed of 30 s-Cepheids according to this criterion.

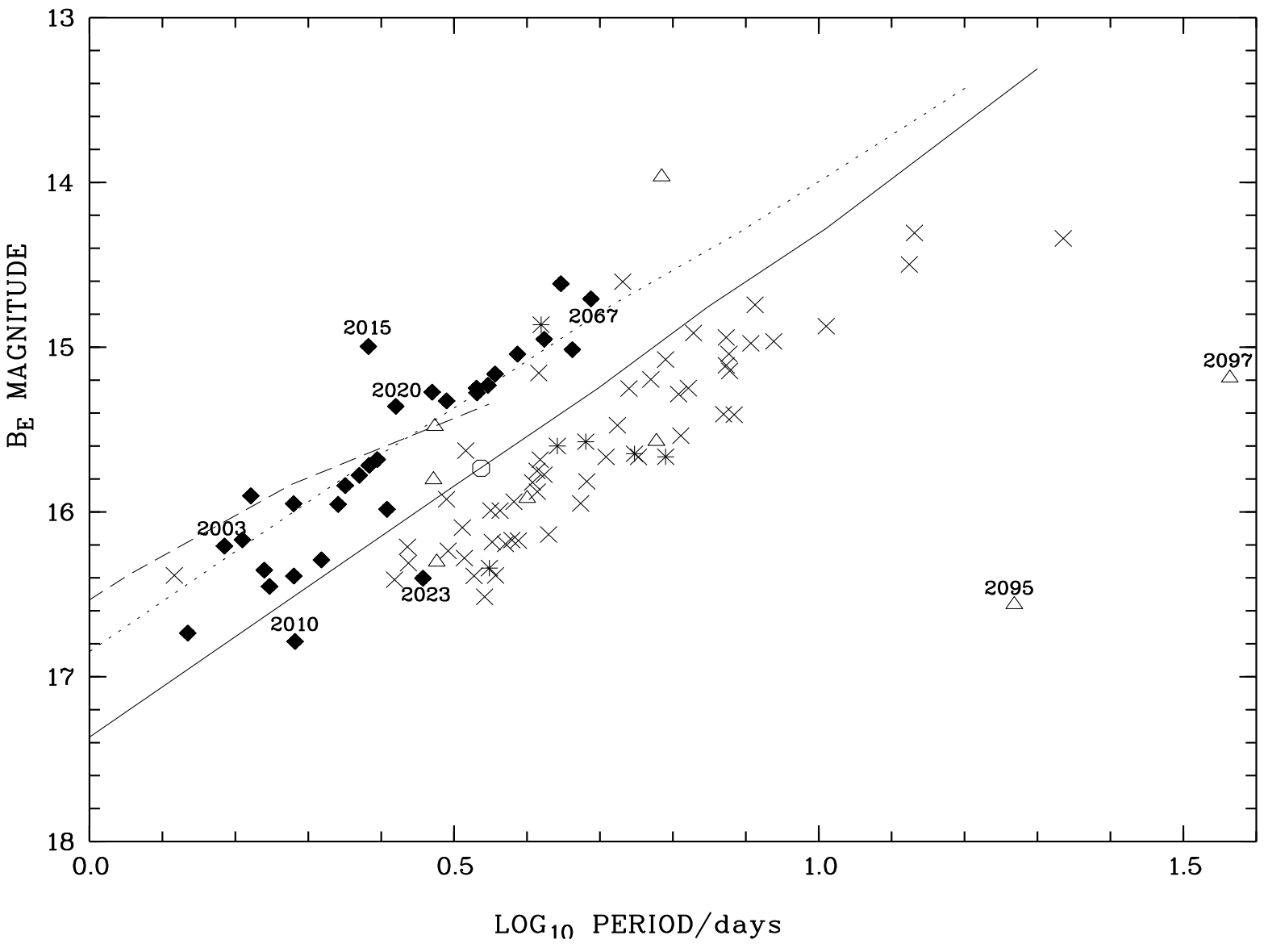

**Fig. 4.** Period-luminosity relation for the 97 EROS Cepheids in the $B_{\rm E}$ filter (mean wavelength=490 nm; same symbols as Fig. 1). Most s-Cepheids are ~1 mag brighter than classical Cepheids of the same period. The full, dotted and dashed lines represent the blue edges of the instability strip for fundamental, first and second overtone pulsation, as calculated by Chiosi et al. (1993) with no convective overshoot.

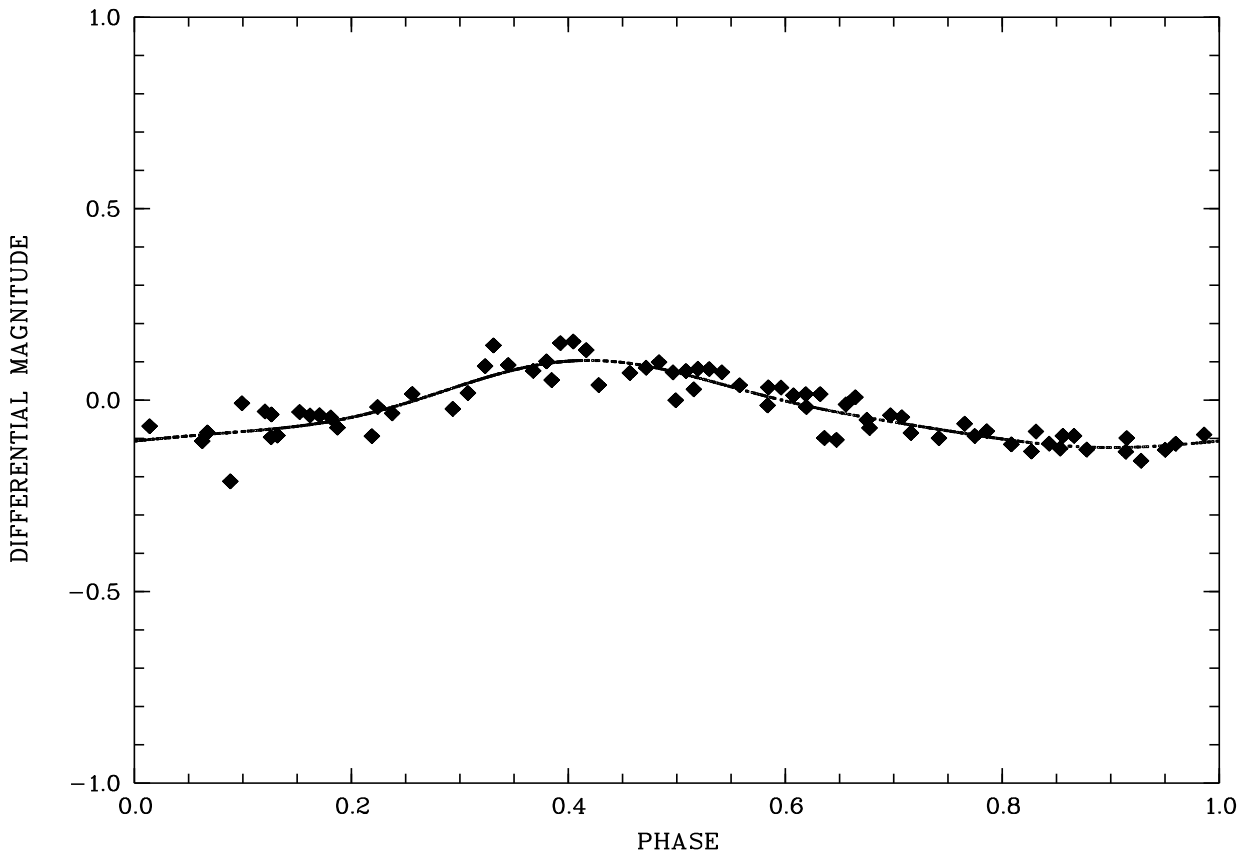

**Fig. 3.** (b) The overtone mode for EROS 2037 plotted at the same vertical scale as Fig. 3a. The points represent the mean $B_{\rm E}$ light curve in 0.01 bins of overtone-mode phase after subtraction of the Fourier model obtained for the fundamental mode. The zero of phase is arbitrary. The full line represents the Fourier model for the overtone mode taken to the third order. The ratio of periods $P(1{\rm H})/P(1{\rm F}) = 0.710 \pm 0.001$.

A group of 6 stars with $R_{21} \approx 0.24$–$0.27$ and $4 < P < 6.5$ d (2040, 2057, 2061, 2065, 2072 and 2078) stands apart as of uncertain appartenance in Figure 5. They are plotted as asterisks.

We classify the remaining 51 stars as classical Cepheids and plot them as diagonal crosses.

## 5. Discussion

### *5.1. Pulsation modes*

For the vast majority of stars, we see in Figure 4 that our *morphological* classification based on Fourier parameters is mirrored by a clear separation into two period-luminosity sequences, with the s-Cepheids ~1.0 mag brighter than the classical ones at any given period. Figure 6 shows the classification is further mirrored by a separation in the $\Phi_{21} - P$ diagram, and we shall see below that this is true of other Fourier parameters too.

An obvious interpretation is that the lower sequence in Figure 4 corresponds to fundamental-mode pulsators and most classical Cepheids (as we define them) are pulsating in this mode, while the upper sequence corresponds to first-overtone pulsators and most of our s-Cepheids pulsate in this mode. Figure 1 then shows that the overtone pulsators are in general bluer, as suspected for Galactic Cepheids by Antonello et al. (1992).

The few stars which deviate from this correspondence clearly merit further observation to determine whether they really do deviate, or whether their EROS photometry is systematically in error, as discussed in Section 2. Examples are EROS 2023, and possibly EROS 2010, which have s-Cepheid values of $R_{21}$ but fundamental-mode values of luminosity and colour, and EROS 2033, 2055 and 2070 where classical Cepheids appear to have first-overtone luminosity and colour. Examination of these latter three stars' images shows that they are oval. Thus they have al-most certainly been brightened by optical binarity, which could also explain their bluer colours.

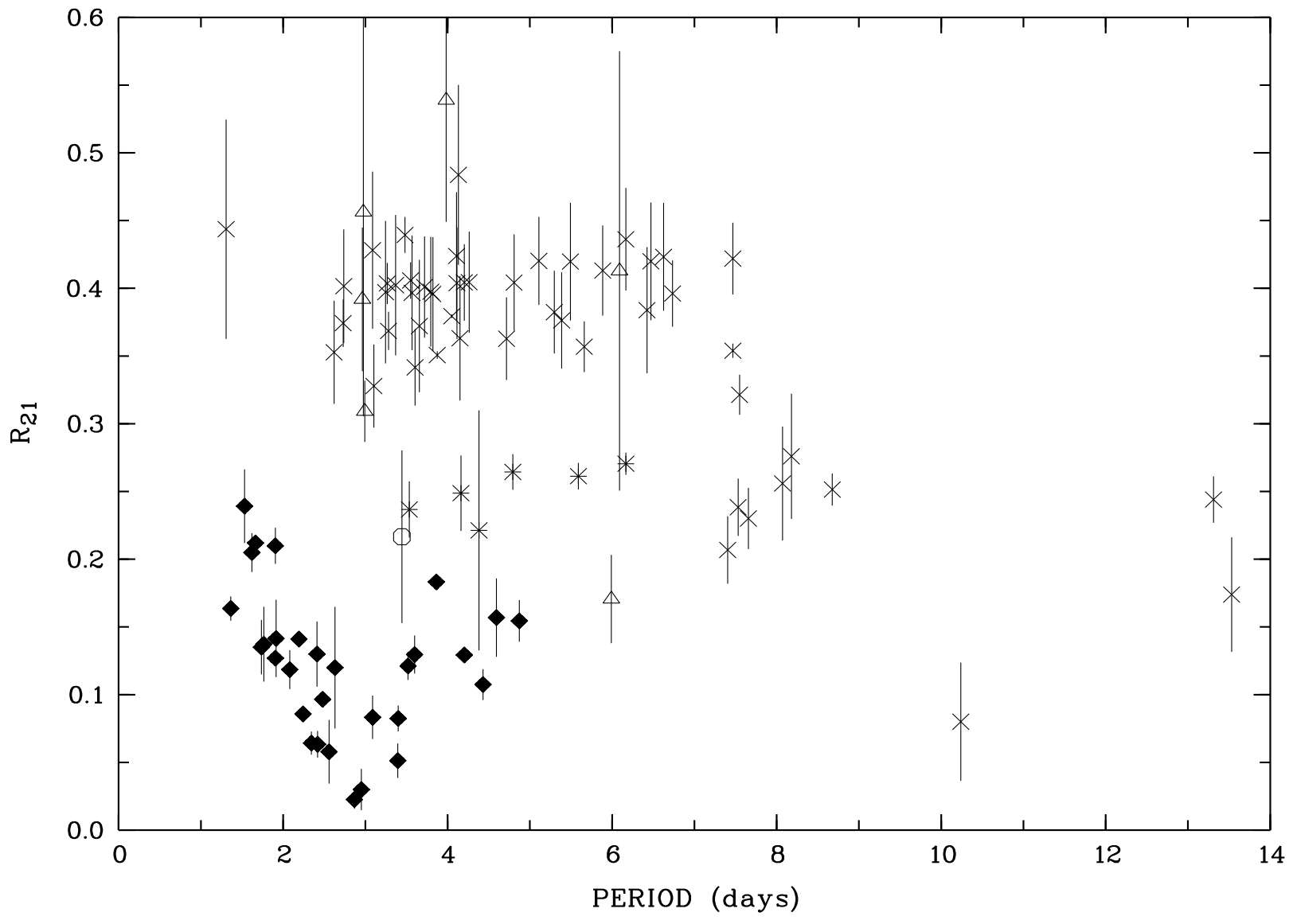

**Fig. 5.** The $R_{21} - P$ diagram for $B_{\rm E}$ light curves for the 94 objects with $P < 14$ d (same symbols as Fig. 1). We define as *s-Cepheids* those non-anomalous stars within the region defined by $R_{21} < 0.30$ for $P < 3$ d and $R_{21} < 0.20$ for $3.0 < P < 5.5$ d.

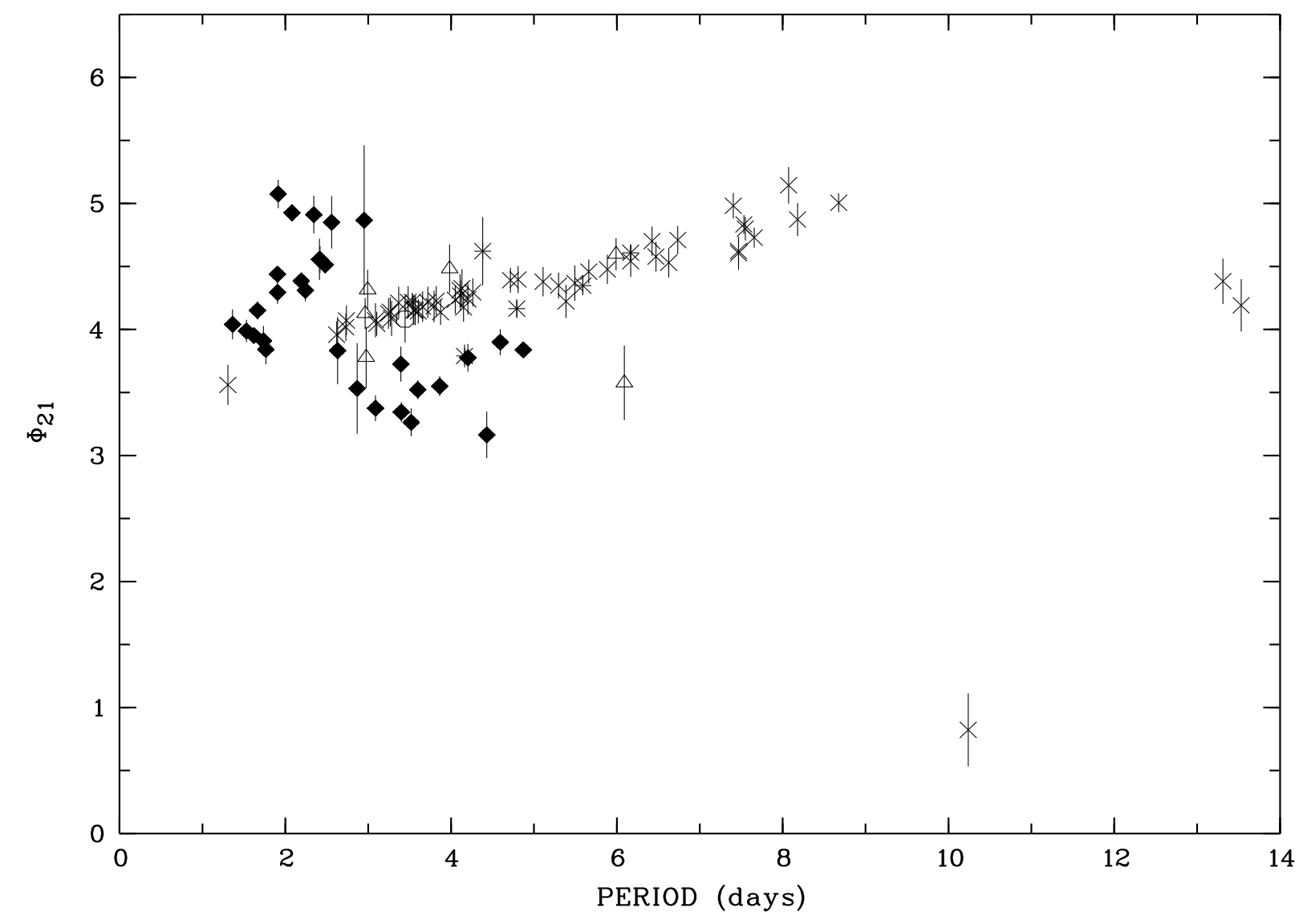

**Fig. 6.** The $\Phi_{21} - P$ diagram for $B_{\rm E}$ light curves for the 94 objects with $P < 14$ d. The classical and s-Cepheids follow different loci.

Figure 4 indicates that the intermediate or indeterminate stars with $R_{21} \sim 0.24$–$0.27$ are a mixed bag: five are fundamental-mode pulsators, one (2057) is an overtone pulsator, and one (2037) is pulsating in both modes simultaneously. Figure 7 confirms that in general the s-Cepheids have lower amplitudes than the classical ones and shows that the intermediate stars, besides having intermediate values of $R_{21}$, have intermediate amplitudes.

The identification of s-Cepheids as overtone pulsators is supported by the theoretical models of Chiosi et al. (1993). In Figures 1 and 4 we have drawn in the blue edges of the fundamental and first and second overtone instability strips transformed from models with no convective overshoot (which Chiosi et al. claim best-matches LMC Cepheids), $Y$=0.25, $Z$=0.008 and the Padova scale. In both figures the fundamental-mode blue edge delimits the distribution of classical Cepheids well. The double-mode Cepheid EROS 2037 even falls close to the fundamental-mode blue edge. However the EROS observations show that first-harmonic blue edge is both brighter and bluer than the Chiosi et al. predictions by $> 0.2\,\mathrm{mag}$.

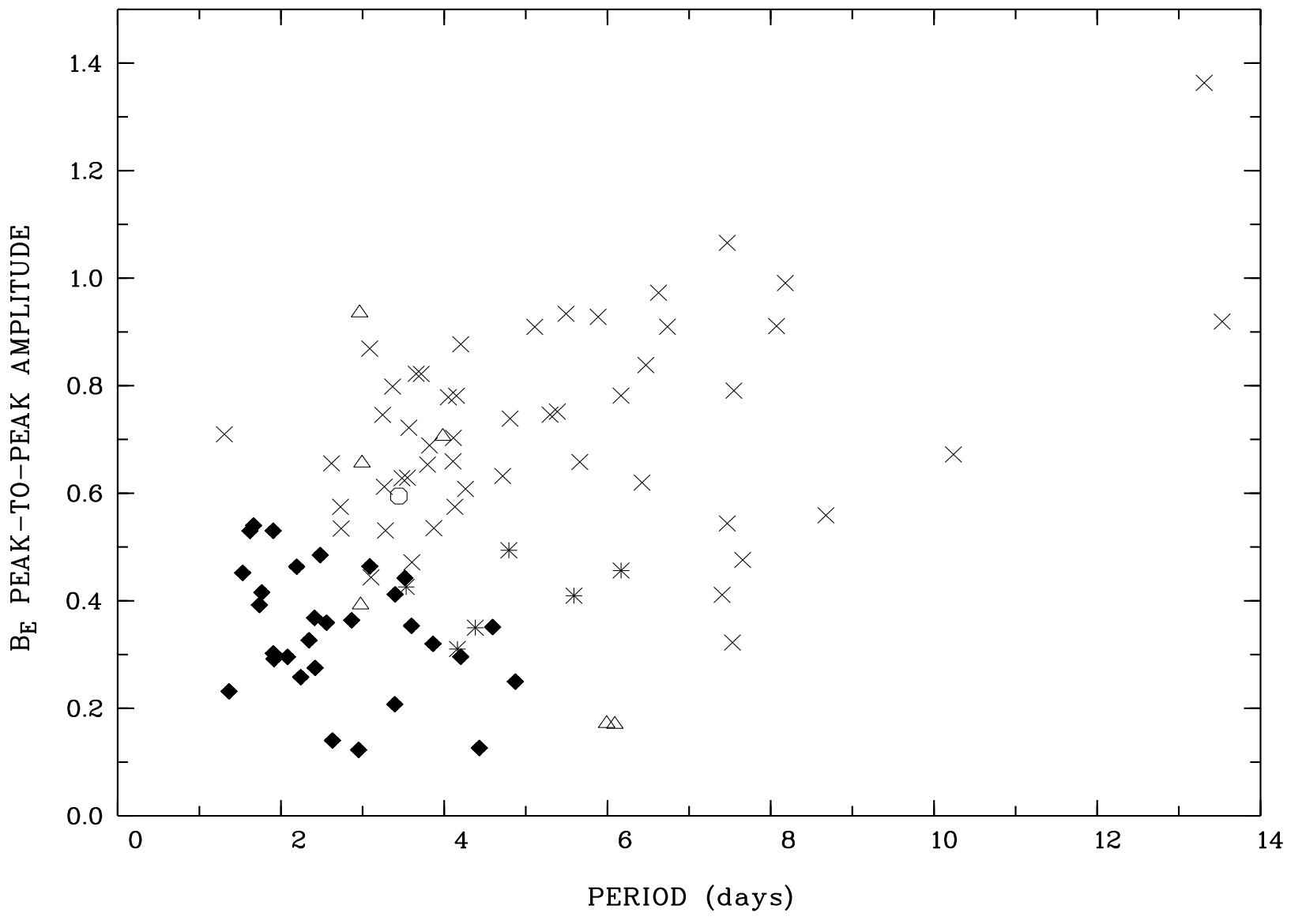

**Fig. 7.** Amplitude-period diagram for the 94 objects with $P < 14\,\mathrm{d}$. The s-Cepheids have a strong tendency to lower amplitudes than the classical Cepheids.

More-detailed comparisons are hardly justified given the zero-point uncertainties in the EROS photometry, but we note that Chiosi et al. models with mild, and especially full overshoot (not plotted) bound the observed distribution of classical Cepheids in Figure 4 less well.

### *5.2. 'Modified period'-Luminosity relation.*

We have plotted the Figure 4 period-luminosity relation in a different way in Figure 8. We have modified the period of the presumed overtone pulsators to the corresponding fundamental period, using the ratio $P(1\mathrm{H})/P(1\mathrm{F}) \approx 0.70$. The classical and s-Cepheid sequences are now closer, but not coincident. This is because a fundamental and first-overtone Cepheid of the same magnitude are not otherwise-identical stars differing only in period: they also differ in colour, as seen in Figure 1.

### *5.3. 'Modified period'-Luminosity-Colour relation.*

Sandage (1958) showed that Cepheids are more accurately represented by a period-luminosity-colour relation. Numerous attempts have been made to determine the relation. It is not critical which one we use: we have chosen the period-luminosity-colour relation for LMC Cepheids determined by Feast (1984):

$$<M_V>= \alpha \log(P) + \beta(<B_0> - <V_0>) + \phi \qquad (6)$$

with $\alpha$=−3.8, $\beta$=2.70 and $\phi$=−2.39.

In Figure 9 we take account of the colour term. The period plotted for a given star was determined in the same manner as in Figure 8. The ordinate is $<V_\mathrm{J}> -\beta(<B_\mathrm{J}> - <V_\mathrm{J}>) +$ const. obtained using Eqs. 4 and 5. If the PLC relation is applicable, the expected slope in Figure 9 is $\alpha$.

The diagram has considerably lower scatter than Figures 4 and 8, and we remark that most of the classical and s-Cepheids lie on the same sequence. *The colour term in the PLC relation is clearly justified, and the same colour correction applies to first-overtone and to fundamental pulsators.*

Some stars do not follow the relation. This may be due to optical or physical binarity. Another possibility is that the star is a second-overtone pulsator (e.g. 2001, 2003, 2015, 2067).

### *5.4. Fourier parameters*

#### 5.4.1. The $\Phi_{21} - P$ plane: fundamental pulsators

The classical Cepheids, or fundamental pulsators, fall along a strip corresponding to the Hertzprung progression. From comparison with Galactic results (Simon 1988, and references therein ) we would expect $\Phi_{21}$ to rise sharply to $2\pi$ at $P \approx 10\,\mathrm{d}$ and thereafter rise from zero towards a constant value for longer periods. This steep change has been interpreted as a 2:1 resonance between the second overtone and the fundamental in the normal mode spectrum of these stars (Simon 1988, and references therein). Andreasen & Petersen (1987) and Andreasen (1988) have shown that in both the LMC and the SMC the resonance occurs around 10 days. From our sample we can conclude that the resonance occurs between 8.7 d (lower limit given by EROS 2091) and 10.2 d (upper limit given by EROS 2092). A better value will be given by a bigger sample of LMC Cepheids.

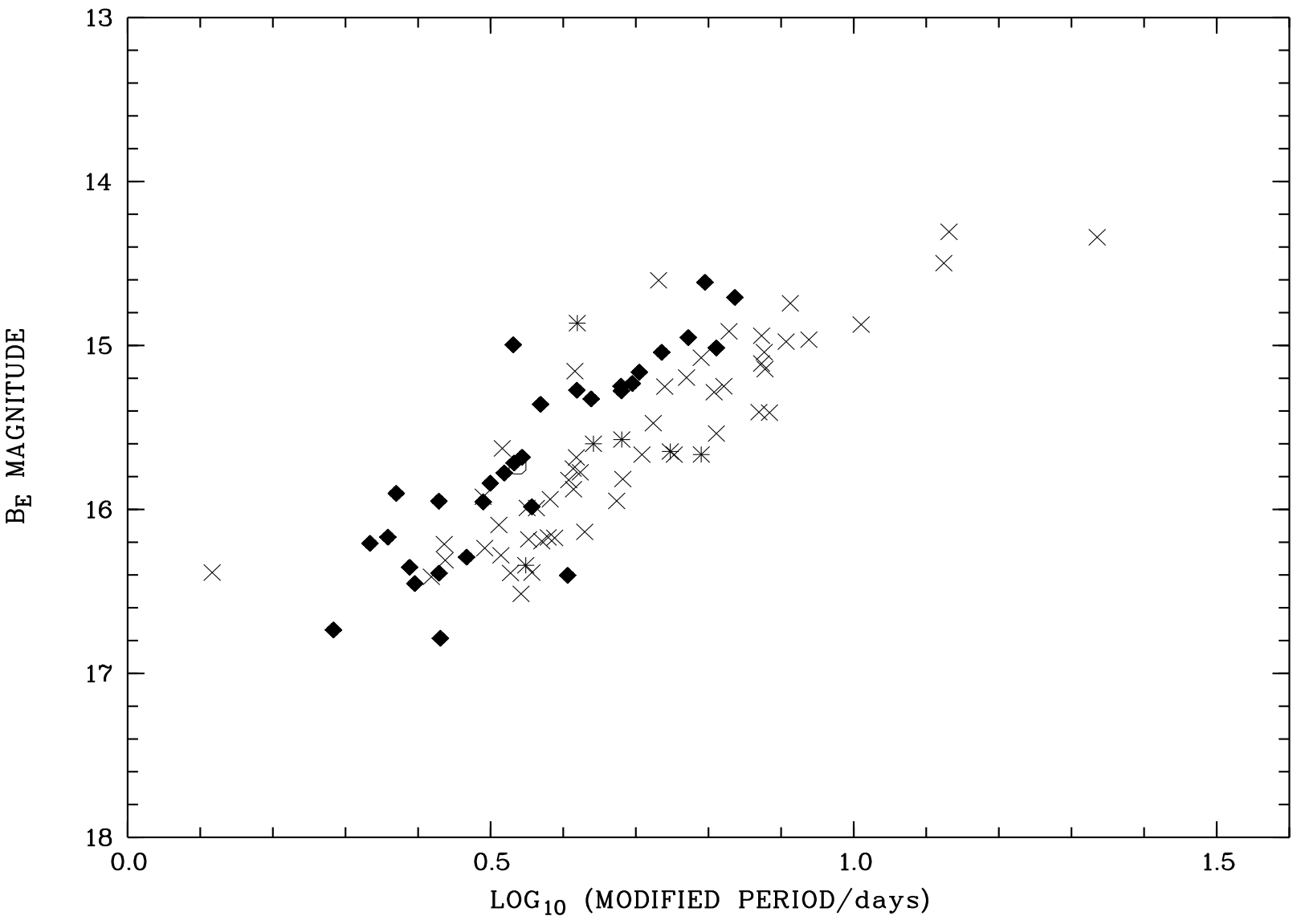

**Fig. 8.** Period-luminosity relation. The periods used in the diagram for the first-overtone candidates and first-overtone blue edge were modified using the relation $P/0.70$. Compare with Fig. 4.

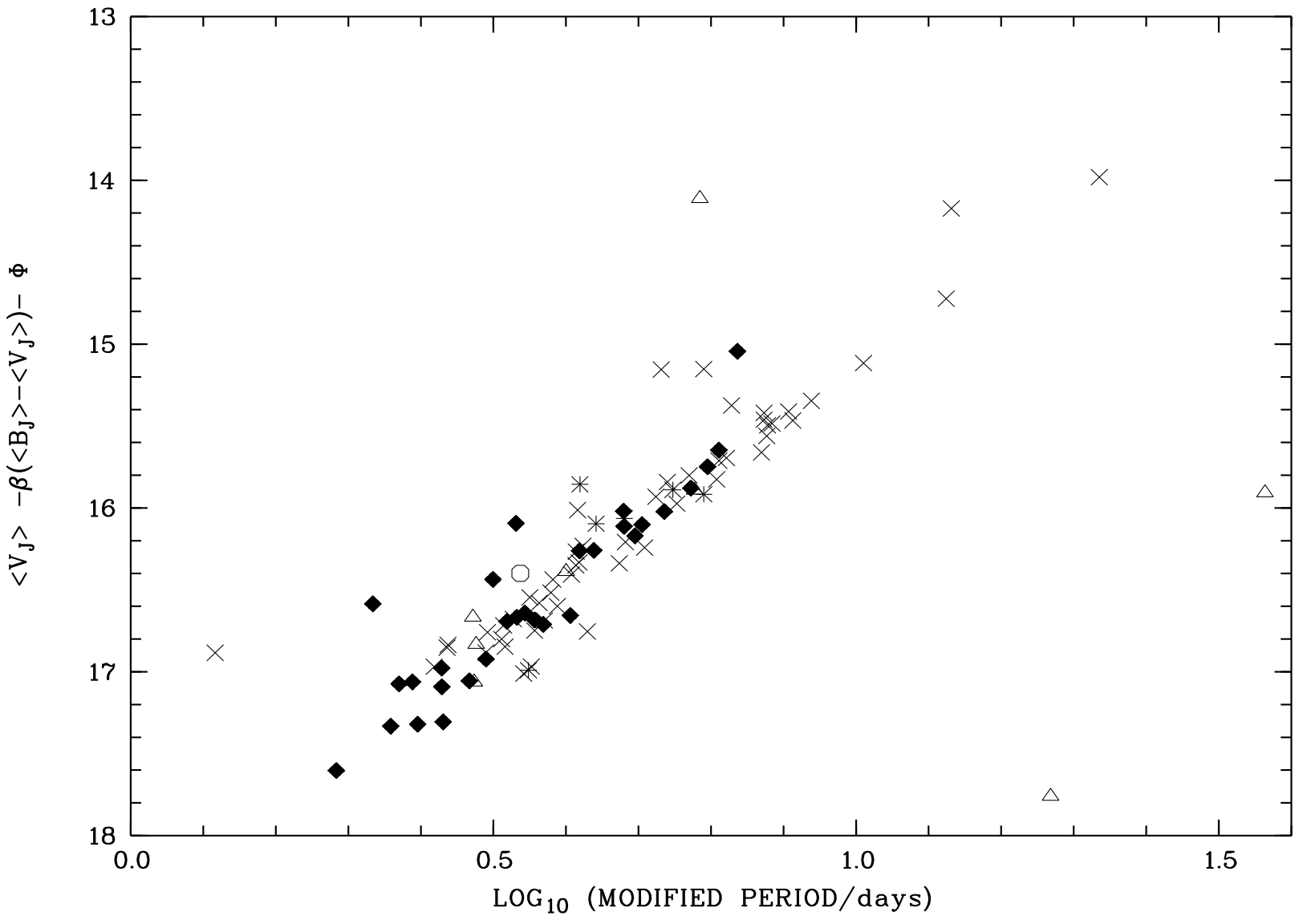

**Fig. 9.** Period-luminosity-colour relation. The periods were modified as in Figure 8. The ordinate is the numerical value $< B_{\rm E} > -2.95(< B_{\rm E} > - < R_{\rm E} >) + 2.39$. Compare with Figures 4 and 8. The classical and s-Cepheids are seen to obey the same PLC relation for the fundamental period.

For LMC Cepheids with periods greater than 12 days, EROS 2093, 2094, and 2096 suggest that the limiting value of $\Phi_{21}$ is about 4 rad, which is similar to the values found by Andreasen & Petersen (1987) for 7 stars with periods greater than 12 days observed photographically.

5.4.2. The $\Phi_{21} - P$ plane: overtone pulsators

For overtone pulsators, two well-defined sequences noted by Antonello et al. (1990) appear. The *upper sequence* is characterized by $P < 3.2\,\mathrm{d}$, $\Phi_{21} > 4.2\,\mathrm{rad}$, and for the *lower sequence* $3.0 < P < 6\,\mathrm{d}$ and $\Phi_{21} < 4.0\,\mathrm{rad}$. These upper and the lower sequences, and their possible link is what we called the Z-shape. It is of great importance to notice that the stars on the horizontals of the Z near where they join the diagonal are located at two or more times their formal error from the classical sequence. Therefore the existence of the Z-shape in the $\Phi_{21} - P$ plane is beyond doubt.

From our earlier discussion we can see that the stars belonging to the lower part of the Z, the lower sequence, have high luminosity and are overtone pulsators. Therefore, the alternative hypothesis of Gieren et al. (1990) that these stars could be fundamental pulsators is refuted.

The drop in the Z-shape is observed in the Galaxy at a period around 3.2 d and has been interpreted as the signature of a possible 2:1 resonance between the first and fourth overtone. A comparison of the Z-shapes exhibited in the Galaxy and the LMC gives information about the effect of metallicity on the position of this possible resonance.

The position of the drop is difficult to estimate. In our sample of s-Cepheids, we decided not to take account of EROS 2023, which despite its low $R_{21}$ would appear to be a 1F pulsator, because its magnitude is faint, and EROS 2024, because of the large uncertainties on its $\Phi_{21}$ value.

Visual inspection of the diagram places the drop in the range 2.5–3.4 days. In the $\Phi_{21} - P$ plane, the limit of the lower sequence is given by EROS 2029 ($P = 3.0882\,\mathrm{d}$, $\Phi_{21}$=3.38 rad), and a limit on the upper sequence can be found from EROS 2018 ($P = 2.5582\,\mathrm{d}$, $\Phi_{21}$=4.85 rad). We do not observe a period overlap between the upper and the lower sequence. EROS 2020 ($P$=2.6303 d, $\Phi_{21}$=3.83 rad) falls between these limits, on top of the classical sequence, about 3 standard deviations below the upper sequence. In our opinion, its observed $\Phi_{21}$ value is a signature of the joining branch between the upper and the lower parts of the Z-shape. With only one object in this branch it is not possible to give a certain value for the position of the drop. We propose that it may be at $P \approx 2.7 \pm 0.2\,\mathrm{d}$, that is, at a shorter period than in the Galaxy. Buchler & Moskalik (1995) observed that in the SMC the drop should occur at a lower period than in the Milky Way, but were unable to give a precise position for it because they only have a few objects in the lower part of their Z-shape.

The EROS CCD data from the 1992-1994 observational campaigns should double the number of Cepheids and lead to a better definition of the overtone sequence.

5.4.3. The $R_{31} - P$, $\Phi_{31} - P$ and $\Phi_{41} - P$ planes.

In Figures 10, 11 and 12 we present $R_{31} - P$, $\Phi_{31} - P$ and $\Phi_{41} - P$ diagrams. Classical-Cepheid features in these planes have already been discussed by earlier authors. Simon & Moffet (1985) have shown that some resonances may present stronger signatures using $\Phi_{31}$ and $\Phi_{41}$ rather than only $\Phi_{21}$. A sharp dip feature appears in $R_{21} - P$ at $P \approx 7.5\,\mathrm{d}$, though its appearance is confused by the slower fall of $R_{21}$ towards the minimum at $P \sim 10$ d. The dip is possibly weakly visible in $\Phi_{21}$. By going to upper harmonics in the $\Phi_{31} - P$ and $\Phi_{41} - P$ planes, the dip appears more and more clearly. This sharp dip involves five EROS stars (2083–2088). It is centered at $7.5 \pm 0.2$ days.

Moskalik et al. (1992) used the Iglesias-Rogers opacities in order to study the beat and bump Cepheid mass discrepancy via non-linear calculations. They found a dip in the $R_{31} - P$ plane at 7.7 d and related it to the sharp feature observed (but not interpreted) by Simon & Moffet (1983) in the $\Phi_{41} - P$ plane near 7 days. Moskalik et al. consider that it can possibly be related to the near 3:1 resonance between the fundamental and the fourth overtone. We conclude that our dip is a possible signature of this 3:1 resonance observed for the first time in the LMC, at the same period as the feature observed in the models of Moskalik et al.

## 6. Conclusion

We have Fourier analyzed the light curves of Cepheids in the bar of the LMC using photometry obtained by the EROS project. This is the first study of a large sample of Cepheids in the LMC using high-quality photometry acquired over a long time scale with a large CCD mosaic. Good phase coverage was obtained for almost all the objects.

Using these light curves we studied Cepheid pulsations in a nearby galaxy with a metallicity different from our own. Because of the low differential reddening and known distance of the LMC, differences in apparent magnitude correspond to differences in absolute magnitude, and it was possible to derive new and convincing results concerning fundamental and overtone pulsations previously only suggested by comparable studies in the Milky Way by Poretti (1994), Simon (1988) and others.

The colour-magnitude diagram was used to select 97 Cepheid or cepheid-like variable stars. Only 29 of them were previously recognised in Payne-Gaposchkin's (1971) Harvard Catalogue.

Eight of the stars present unusual light curves, and one is a double-mode Cepheid. Almost all the rest divide cleanly into two groups in the $R_{21} - P$ diagram. We associate the group with periods less than about 5.5 d and low values of $R_{21}$ with the s-Cepheids and the rest with classical Cepheids, though this definition is not of course

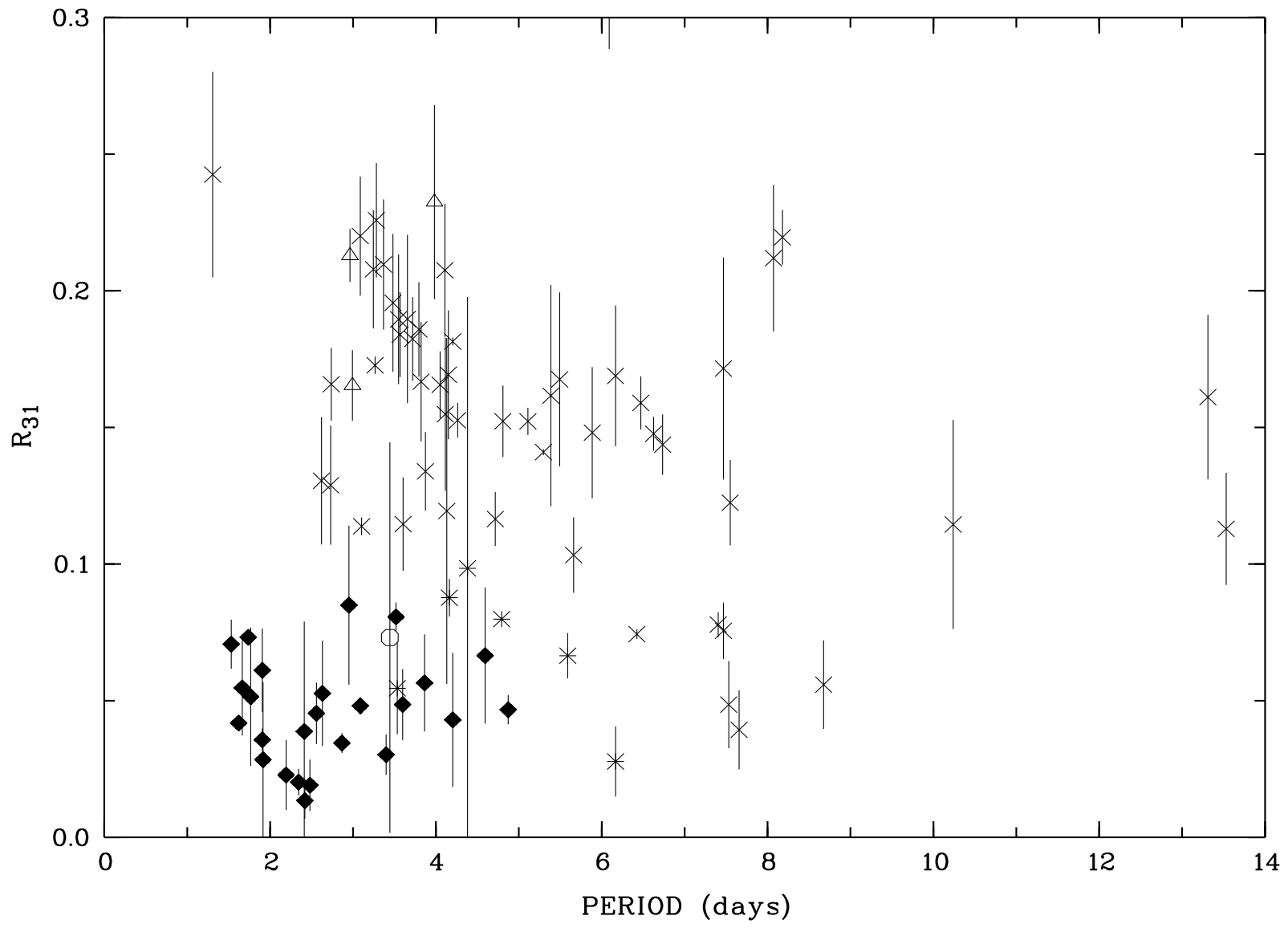

**Fig. 10.** The $R_{31} - P$ diagram for $B_{\rm E}$ light curves for the 94 objects with $P < 14$ d. The sharp dip for classical Cepheids at $P \approx 7.5$ d is stronger than in Fig. 6.

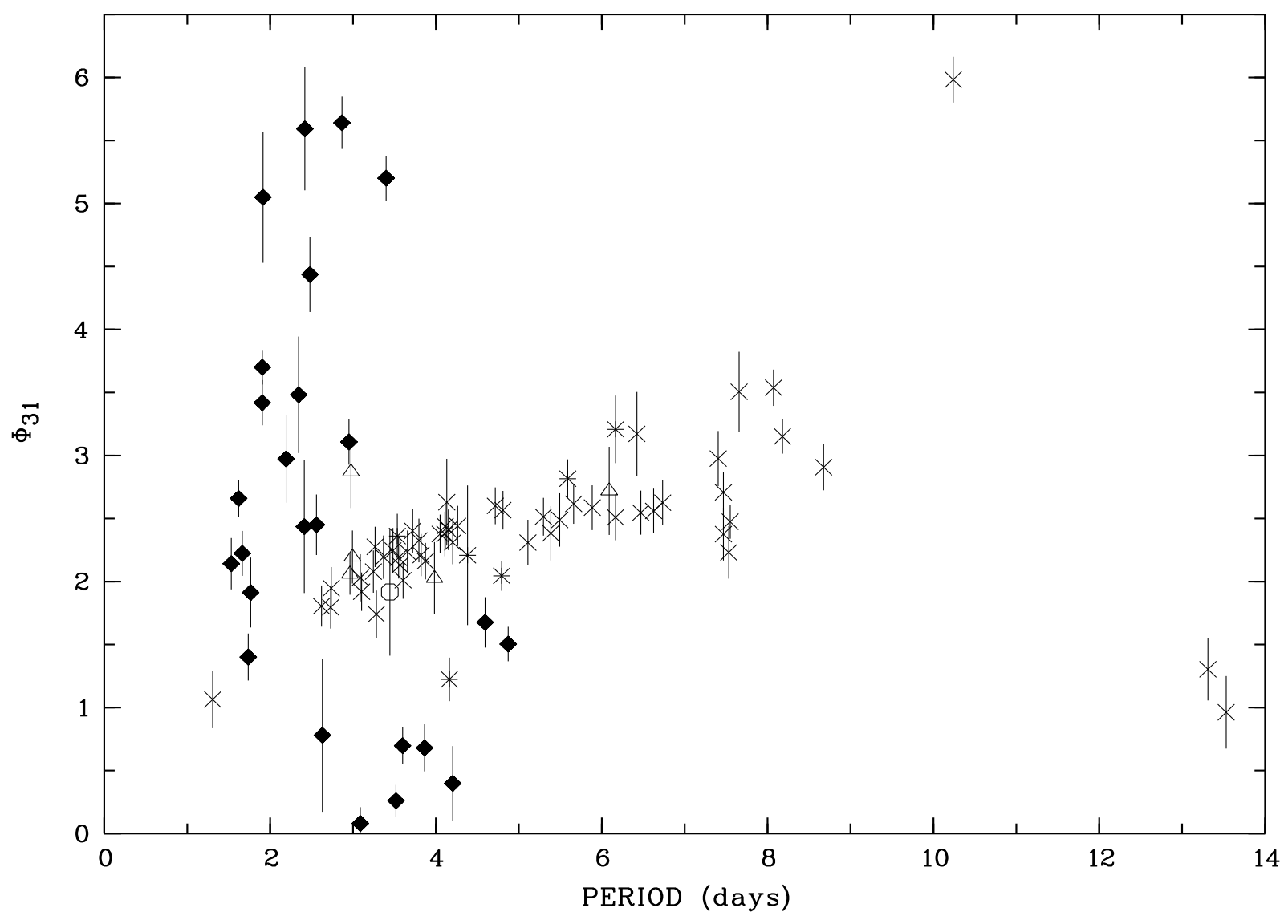

**Fig. 11.** The $\Phi_{31} - P$ diagram for $B_{\rm E}$ light curves for the 94 objects with $P < 14$ d. The sharp dip for classical Cepheids at $P \approx 7.5$ d is stronger than in Fig. 6.

phenomenologically identical to that adopted by other authors.

The stars in the two groups mostly follow separate, parallel loci in the period-luminosity diagram, with the s-Cepheids being more luminous by $\sim$1 mag for any given period. It has been suggested that s-Cepheids are first-overtone pulsators, and theoretically it is expected that the ratio of periods $P(1\rm H)/P(1\rm F) \sim 0.7$. When the s-Cepheid periods are modified by this factor and a standard LMC period-colour-luminosity relation is applied, the classical and s-Cepheid loci are found to be conincident, furnishing persuasive evidence that s-Cepheids are pulsating in the first-overtone and that the colour term is justified in the PLC relation. Further, the overtone pulsators are found in the colour-magnitude diagram to occupy the bluer portion of the instability strip, indicating that it is colour which determines whether a Cepheid pulsates in its fundamental or first harmonic. The blue edge of the fundamental-mode instability strip agrees well with the theoretical calculations of Chiosi et al. (1993), but the first-overtone blue edge is observed to be bluer and brighter by $> 0.2$ mag than the Chiosi et al. predictions.

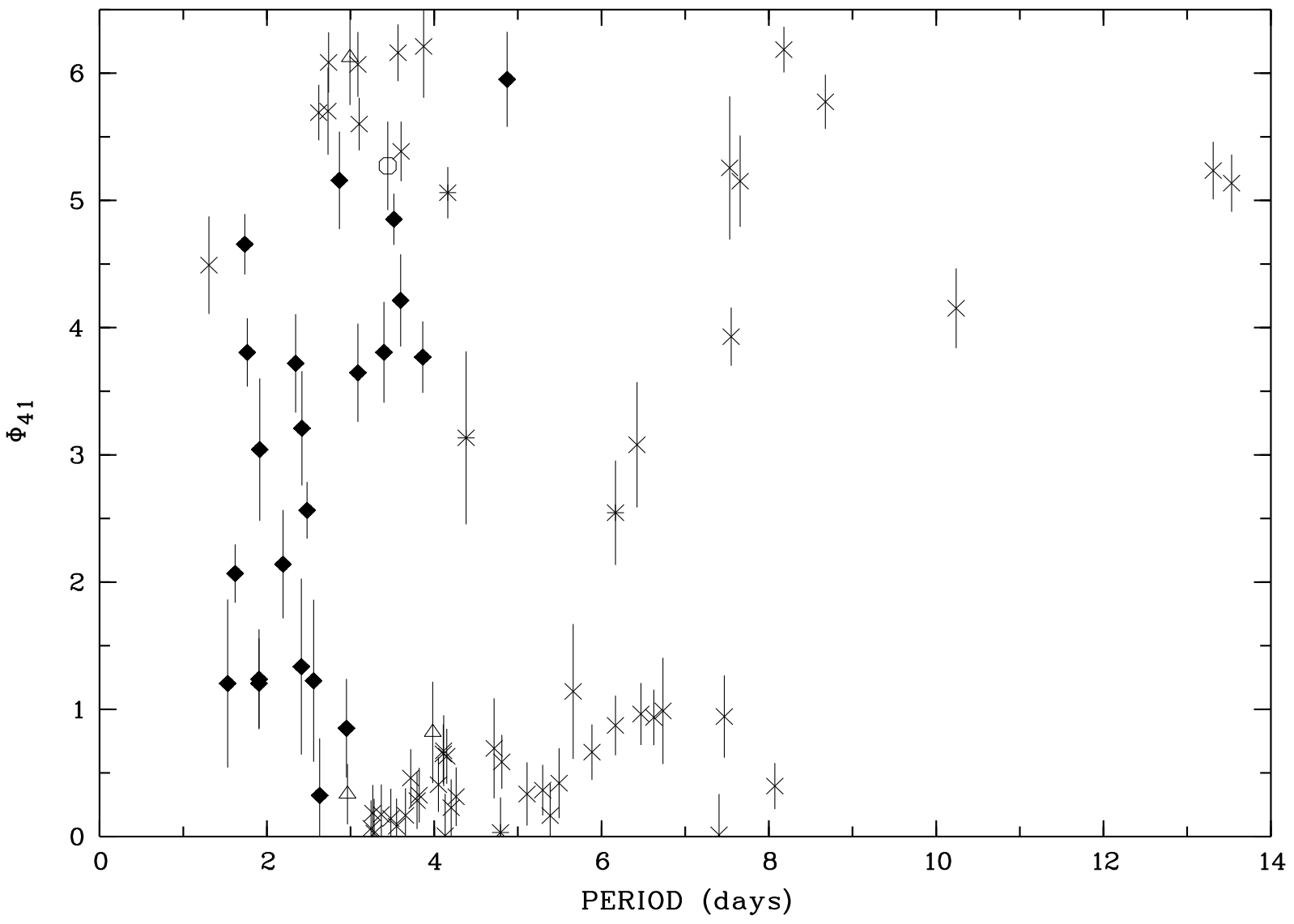

**Fig. 12.** The $\Phi_{41} - P$ diagram for $B_E$ light curves for the 94 objects with $P < 14$ days. The sharp dip for classical Cepheids at $P \approx 7.5$ d is stronger than in Figures 6 and 11.

The overtone pulsators also tend to have smaller amplitudes and are separated from fundamental pulsators in the $\Phi_{21} - P$ diagram.

Seven stars have intermediate values of $R_{21}$ and light-curve amplitude. They prove to be of varied nature: 5 fundamental-mode pulsators, one overtone pulsator and one double-mode Cepheid.

We identify about 50 fundamental and 30 obvious overtone pulsators, and most of these stars have periods less than 10 d. Payne-Gaposchkin & Gaposchkin (1971) showed that the LMC was overabundant in short-period Cepheids compared to the Galaxy, but detected only 3% of s-Cepheids. This low proportion undoubtedly reflects the difficulty of detecting low amplitudes in photographic data. We find, as suspected by Matteo et al. (1990) on the basis of a sample of 6 stars, that the proportion of LMC overtone pulsators is much higher, representing ~30% of our sample, and ~40% of stars with $P < 5.5$ d.

Some Cepheids are brighter than expected from the modified period-colour-luminosity relation. It is known that the EROS photometry can be uncertain for individual stars, but the deviant objects might also be Cepheids with optical or physical binary companions, or possibly cases of second overtone pulsators. These possibilities could be tested with better-quality photometry.

Other stars requiring further photometry include EROS 2023, and possibly EROS 2010, which have sinusoidal light curves but classical-Cepheid luminosities. If these luminosities are confirmed, it will indicate that 1F pulsators can occasionally have symmetric, low-amplitude light curves.

The features exhibited in the $R_{21} - P$ and $\Phi_{21} - P$ planes are similar to those found for Galactic Cepheids. Most of our stars have period less than 10 days: however, it is possible to constrain the position of the 2:1 resonance between the fundamental and second overtone for classical Cepheids. We think that this resonance occurs between 8.7 and 10 d, as already suggested by Andreasen & Petersen (1987).

A sharp feature near 7.5 d is observed in the Fourier planes (especially the higher-order ones). We suggest it is related to a similar feature observed in the models of Moskalik et al. (1992), which in turn is believed related to the near-3:1 resonance between fundamental and fourth overtone.

The Z-shape around ~3 days in the $\Phi_{21} - P$ diagram is clearly seen. The overtone nature of s-Cepheids with period greater than 3 days is confirmed by their higher luminosity than classical Cepheids. Moreover, the 'drop' of the Z-shape for LMC first-overtone pulsators is estimated to be at $\approx 2.7 \pm 0.2$ d. This value is smaller than the one observed for Galactic Cepheids. This may be due to the lower metallicity of the LMC.

We thus find that the position of resonances in overtone pulsators is shifted in the LMC, whereas for fundamental pulsators the values are similar to Galactic ones.

Using the 1993-1994 EROS CCD data, we expect to double the number of Cepheids in our sample. This will better define the positions of the resonances. Moreover, comparison with observations of similar nature carried out in the SMC, like the work of Buchler & Moskalik (1994), should be of great interest.

*Acknowledgements.* We thank Michael Feast, Marie Jo Goupil, Douglas Welch and Christophe Alard for very helpful discussions. J.D.P. thanks the IAU Travel Grant Scheme, the Frank Bradshaw and Elizabeth Pepper Wood Fund, the Prince & Princess of Wales Science Award and the Kingdon-Tomlinson Trust for financial support which enabled him to work with the EROS project in Paris.

## Appendix. Uncertainties for the Fourier decomposition parameters

For a model of the form given by Eq. (1), uncertainties for the Fourier decomposition parameters $R_{k1}$ and $\Phi_{k1}$ have already been calculated and discussed by Petersen (1986). In this paper we take a different model, because the Generalized Periodogram method introduced by Grison (1994) requires a different form of Fourier decomposition.

Grison's procedure is to take a model for a curve of zero mean and of form

$$\sum_{i=1}^{M}(A_i \cos(i\frac{2\pi}{P}t) + B_i \sin(i\frac{2\pi}{P}t)) \qquad (7)$$

where $A_i$ and $B_i$ can be expressed by the relations :

$$A_i = a_i - \sum_{j=i+1}^{M} (\alpha^c_{ij} a_j + \alpha^s_{ij} b_j)$$

$$B_i = b_i - \sum_{j=i+1}^{M} (\beta^c_{ij} a_j + \beta^s_{ij} b_j).$$

$R_{k1}$ can be expressed as a function of $A_i$ and $B_i$ :

$$R^2_{k1} = \frac{(A^2_k + B^2_k)}{(A^2_1 + B^2_1)}.$$

The $a_j, b_j, \alpha^c_{ij}, \beta^c_{ij}\alpha^s_{ij}, \beta^s_{ij}$ are determined by the Generalized Periodogram method. The $a_j$ and $b_j$ are random independent variables. Thus, they are uncorrelated. Therefore, by approximating the function $R_{k1} = f(a_i, b_j)$ by its first-order Taylor development around $< R_{k1} >$, the variance $\sigma^2_{R_{k1}}$ can be given by the expression:

$$\sigma^2_{R_{k1}} = \sum_{i=1}^{M}\left(\frac{\partial R_{k1}}{\partial a_i}\right)^2 \sigma^2_{a_i} + \left(\frac{\partial R_{k1}}{\partial b_i}\right)^2 \sigma^2_{b_i} \qquad (8)$$

where

$$\left(\frac{\partial R_{k1}}{\partial a_i}\right) = \frac{1}{2}\left(f_{R_{k1}}\frac{\partial N_{R_{k1}}}{\partial a_i} - g_{R_{k1}}\frac{\partial D_{R_{k1}}}{\partial a_i}\right)$$

$$\left(\frac{\partial R_{k1}}{\partial b_i}\right) = \frac{1}{2}\left(f_{R_{k1}}\frac{\partial N_{R_{k1}}}{\partial b_i} - g_{R_{k1}}\frac{\partial D_{R_{k1}}}{\partial b_i}\right)$$

with

$$N_{R_{k1}} = (A^2_k + B^2_k)$$

$$D_{R_{k1}} = (A^2_1 + B^2_1)$$

$$f_{R_{k1}} = \left(\frac{D_{R_{k1}}}{N_{R_{k1}}}\right)^{\frac{1}{2}} \frac{1}{D_{R_{k1}}} = R_{k1}\frac{1}{D_{R_{k1}}}$$

$$g_{R_{k1}} = \left(\frac{N_{R_{k1}}}{D_{R_{k1}}}\right)^{\frac{1}{2}} \frac{1}{D_{R_{k1}}} = \frac{1}{R_{k1}}\frac{1}{D_{R_{k1}}}$$

and

$$\frac{\partial N_{R_{k1}}}{\partial a_i} = \begin{cases} 0 & \text{i < k} \\ 2A_k & \text{i = k} \\ -2(\alpha^c_{ki}A_k + \beta^c_{ki}B_k) & \text{i > k} \end{cases}$$

$$\frac{\partial N_{R_{k1}}}{\partial b_i} = \begin{cases} 0 & \text{i < k} \\ 2B_k & \text{i = k} \\ -2(\alpha^s_{ki}A_k + \beta^s_{ki}B_k) & \text{i > k} \end{cases}$$

$$\frac{\partial D_{R_{k1}}}{\partial a_i} = \begin{cases} 2A_1 & \text{i = 1} \\ -2(\alpha^c_{1i}A_1 + \beta^c_{1i}B_1) & \text{i > 1} \end{cases}$$

$$\frac{\partial D_{R_{k1}}}{\partial b_i} = \begin{cases} 2B_1 & \text{i = 1} \\ -2(\alpha^s_{1i}A_1 + \beta^s_{1i}B_1) & \text{i > 1.} \end{cases}$$

The same reasoning applies to the phase differences $\Phi_{k1}$ for which it is easy to show that

$$\Phi_i = \arctan\left(\frac{-B_i}{A_i}\right) \quad i = 1, ..., M. \qquad (9)$$

(We define $\Phi_i$ over $[-\pi, \pi]$ ; the sign of $\Phi_i$ is given by the sign of $-B_i$.) Therefore the variance $\sigma^2_{\Phi_{k1}}$ is given by the relation:

$$\sigma^2_{\Phi_{k1}} = \sum_{i=1}^{M}\left(\frac{\partial \Phi_{k1}}{\partial a_i}\right)^2 \sigma^2_{a_i} + \left(\frac{\partial \Phi_{k1}}{\partial b_i}\right)^2 \sigma^2_{b_i} \qquad (10)$$

where

$$\frac{\partial \Phi_{k1}}{\partial a_i} = \frac{\partial \Phi_k}{\partial a_i} - k\frac{\partial \Phi_1}{\partial a_i}$$

$$\frac{\partial \Phi_{k1}}{\partial b_i} = \frac{\partial \Phi_k}{\partial b_i} - k\frac{\partial \Phi_1}{\partial b_i}$$

with

$$\frac{\partial \Phi_k}{\partial a_i} = \begin{cases} 0 & \text{i < k} \\ \frac{h_k B_k}{A^2_k} & \text{i = k} \\ \frac{h_k}{A^2_k}(-\alpha^c_{ki}B_k + \beta^c_{ki}A_k) & \text{i > k} \end{cases}$$

$$\frac{\partial \Phi_1}{\partial a_i} = \begin{cases} \frac{h_1 B_1}{A^2_1} & \text{i = 1} \\ \frac{h_1}{A^2_1}(-\alpha^c_{1i}B_1 + \beta^c_{1i}A_1) & \text{i > 1} \end{cases}$$

$$\frac{\partial \Phi_k}{\partial b_i} = \begin{cases} 0 & \text{i < k} \\ -\frac{h_k}{A_k} & \text{i = k} \\ \frac{h_k}{A^2_k}(-\alpha^s_{ki}B_k + \beta^s_{ki}A_k) & \text{i > k} \end{cases}$$

$$\frac{\partial \Phi_1}{\partial b_i} = \begin{cases} -\frac{h_1}{A_1} & \text{i} = 1 \\ \frac{h_1}{A_1^2}(-\alpha^s_{1i}B_1 + \beta^s_{1i}A_1) & \text{i} > 1. \end{cases}$$

and

$$h_k = \frac{1}{1 + (\frac{B_k}{A_k})^2}.$$

## References

Alcock C., Allsman R.A., Axelrod T.S. et al., 1995, AJ in press
Andreasen G.K., 1988, A&A 196, 159
Andreasen G.K., Petersen J.O., 1987, A&A 180, 129
Antonello E., Aikawa T., 1993, A&A 279, 119
Antonello E., Poretti E., 1986, A&A 169, 149
Antonello, E., Poretti, E., Reduzzi, L., 1990, A&A 236, 138
Antonello E., Poretti E., Reduzzi L., 1992, Mem. Soc. Astron. Ital. 63, 75
Arnaud, M. Aubourg E., Bareyre P., et al., 1994a, Experimental Astron. 4, 265
Arnaud, M. Aubourg E., Bareyre P., et al., 1994b, Experimental Astron. 4, 279
Arp, H.C., 1960, AJ 65, 404
Aubourg E., Bareyre P., Brehin S., et al., 1993a, The Messenger 72, 20
Aubourg E., Bareyre P., Brehin S., et al., 1993b, Nat 365, 623
Balona L.A., 1984, The double-mode Cepheids. In: Madore, B.F. (ed) Proc. IAU Colloq. 82, Cepheids, theory and observations. Cambridge University Press, Cambridge p.17
Buchler J.R., Moskalik P., 1994, A&A 292, 450
Chiosi, C., Wood, P.R., Capitano, N., 1993, ApJS 86, 541
Feast, M.W., 1984, Magellanic Cloud Cepheids. In: van den Bergh, S., de Boer, K.S. (eds) Proc. IAU Symp. 108, Structure and evolution of the Magellanic Clouds. Reidel, Dordrecht p.157
Gascoigne, S.C.B., 1960, Trans. IAU 10, 680
Gieren W.P., 1982, PASP 94, 960
Gieren W.P., Moffet T.J., Barnes T.G., Matthews, J.M., Frueh, M.L., 1990, AJ 99, 1196
Grison P., 1994, A&A 289, 404
Grison P., Beaulieu J.P., Pritchard J.D., et al., 1995, A&AS 109, 1
Groth, E.J., 1975, ApJS 29, 285
Hertzsprung, E., 1926, Bull. Astron. Inst. Netherlands 3, 115
Iben, I, Jr., Tuggle, R.S., 1975, ApJ 197, 39
Kholopov P.N., Samus' N.N., Frolov M.S., et al., 1985, General Catalogue of Variable Stars, 4th edition, Vol. I. Nauka, Moscow
Mantegazza I., Poretti E., 1992, A&A 261, 137
Mateo M., Olszewski E.W., Madore B.F., 1990, ApJ 353, L11
Moskalik P., Buchler J.R., Marom M., 1992, ApJ 385, 685
Page, T., Carruthers, G.R., 1981, ApJ 248, 906
Payne-Gaposchkin, C.H., 1961, Vistas Astron. 4, 184
Payne-Gaposchkin, C.H., 1971, Smithson. Contr. Astrophys. No. 13
Payne-Gaposchkin, C.H., 1973, Comparison of the Cepheid Variables in the Magellanic Clouds and the Galaxy. In: Muller, A.B. (ed) The Magellanic Clouds. Reidel, Dordrecht p.34
Payne-Gaposchkin, C.H., Gaposchkin, S., 1966, Vistas Astron. 8, 191
Petersen J.O., 1986, A&A 170, 59
Poretti E., 1994, A&A 285, 524
Sandage A., 1958, ApJ 127, 513
Scargle, J.D., 1982, ApJ 263, 835
Schwering, P.B.W., Isreal, F.P., 1990, Atlas and catalogue of infrared sources in the Magellanic Clouds. Kluwer, Dordrecht
Shapley, H., McKibben Nail, V., 1955, Proc. Nat. Acad. Sci. 41, 185
Simon N.R., 1985, ApJ 299, 723
Simon N.R., 1988, The Fourier decomposition technique: Interpreting the variations of pulsating stars. In: Stalio, R., Willson, L.A. (eds) Pulsation and Mass Loss in Stars. Kluwer, Dordrecht p.27
Simon N.R., Lee A., 1981, ApJ 248, 291
Simon N.R., Moffett, T.J., 1983, PASP 97, 1078
Stobie, R.S., 1977, MNRAS 180, 631
Wayman P.A., Stift M.J., Buchler C.J., 1984, A&AS 56, 169
Westerlund B.E., 1993, The stellar populations in the Magellanic Clouds, an overview. In: Baschek, B., Klare, G., Lequeux, J. (eds) New aspects of Magellanic Cloud Research. Springer-Verlag, Berlin p.7